\documentclass[aps,twocolumn,showpacs,preprintnumbers,floats]{revtex4}\def\@cite#1#2{\textsuperscript{[{#1\if@tempswa , #2\fi}]}}
\usepackage{mathrsfs}
\usepackage{longtable,lscape}
\usepackage{txfonts}
\usepackage{amssymb}
\usepackage{indentfirst}
\usepackage{graphicx,booktabs}
\usepackage{float}

\usepackage{multirow}
\usepackage{color}
\usepackage{epsfig}

\begin{document}

\title{Strong decays of the $1P$ and $2D$ doubly charmed states}

\author{Li-Ye Xiao$^{1,2}$~\footnote {E-mail: lyxiao@pku.edu.cn}, Qi-Fang L\"u$^{3,4}$~\footnote {E-mail: lvqifang@hunnu.edu.cn}, and Shi-Lin Zhu$^{1,2,5}$~\footnote {E-mail: zhusl@pku.edu.cn}}

\affiliation{ 1) School of Physics and State Key Laboratory of
Nuclear Physics and Technology, Peking University, Beijing 100871,
China } \affiliation{ 2)  Center of High Energy Physics, Peking
University, Beijing 100871, China} \affiliation{ 3) Department of
Physics, Hunan Normal University, and Key Laboratory of
Low-Dimensional Quantum Structures and Quantum Control of Ministry
of Education, Changsha 410081, China } \affiliation{ 4) Synergetic
Innovation Center for Quantum Effects and Applications (SICQEA),
Hunan Normal University, Changsha 410081, China} \affiliation{ 5)
Collaborative Innovation Center of Quantum Matter, Beijing 100871,
China}


\begin{abstract}

We perform a systematical investigation of the strong decay
properties of the low-lying $1P$- and $2D$-wave doubly charmed
baryons with the $^3P_0$ quark pair creation model. The main
predictions include: (i) in the $\Xi_{cc}$ and $\Omega_{cc}$ family,
the $1P~\rho$ mode excitations with $J^P=1/2^-$ and $3/2^-$ should
be the fairly narrow states. (ii) For the $1P~\lambda$ mode
excitations, $|^2P_{\lambda}\frac{3}{2}^-\rangle$ and
$|^4P_{\lambda}\frac{3}{2}^-\rangle$ have a width of $\Gamma\sim150$
MeV, and mainly decay into the $J^P=3/2^+$ ground state. Meanwhile,
$|^2P_{\lambda}\frac{1}{2}^-\rangle$ and
$|^4P_{\lambda}\frac{5}{2}^-\rangle$ are the narrow states with a
width of $\Gamma\sim40$ MeV, and mainly decay into the ground state
with $J^P=1/2^+$. (iii) The $2D_{\rho\rho}$ states mainly decay via
emitting a heavy-light meson if their masses are above the threshold
of $\Lambda_cD$ or $\Xi_cD$, respectively. Their strong decay widths
are sensitive to the masses and can reach several tens MeV. (iv) The
$2D_{\lambda\lambda}$ states may be broad states with a width of
$\Gamma>100$ MeV. It should be emphasized that the states with
$J^P=3/2^+$ and $5/2^+$ mainly decay into the ground state with
$J^P=3/2^+$ plus a light-flavor meson, while the states with
$J^P=1/2^+$ and $7/2^+$ mainly decay into the ground state with
$J^P=1/2^+$ plus a light-flavor meson.

\end{abstract}


\maketitle

\section{Introduction}

Fifteen years ago, the SELEX Collaboration announced a doubly
charmed baryon $\Xi^+_{cc}$ with mass 3519$\pm$1
MeV~\cite{Mattson:2002vu}. One year later, another doubly charmed
baryon $\Xi^{++}_{cc}$ was reported at 3770 MeV by the same
collaboration~\cite{Moinester:2002uw}. Unfortunately, those two
signals $\Xi^+_{cc}(3519)$ and $\Xi^{++}_{cc}(3770)$ were not
confirmed by other collaborations. Recently, the LHCb Collaboration
discovered a doubly charmed baryon $\Xi^{++}_{cc}(3621)$ in the
$\Lambda^+_cK^-\pi^+\pi^+$ mass spectrum~\cite{Aaij:2017ueg}. Its
mass was measured to be 3621.40$\pm$0.72$\pm$0.27$\pm0.14$ MeV. The
newly observed $\Xi^{++}_{cc}(3621)$ may provide an access point for
the study of doubly heavy baryons and has attracted significant
attention from the hadron physics community
\cite{n1,n2,Wang:2017mqp,n4,n5,n6,n7,n8,Li:2017,Yu:2017,Hu:2005gf,Hu:2017dzi,Xiao:2017udy,Chen:2017jjn,Chen:2017sbg,Lu:2017meb}.

In the past score years, the properties of the doubly heavy baryons
were extensively explored with various theoretical methods and
models including the mass
spectra~\cite{Roncaglia:1995az,Gershtein:2000nx,Itoh:2000um,Ebert:2002ig,Roberts:2007ni,Yoshida:2015tia,Shah:2016vmd,Gershtein:1998sx,Zhang:2008rt,Wang:2010hs,Brown:2014ena}
and semi-leptonic
decays~\cite{Faessler:2001mr,Albertus:2006ya,Roberts:2008wq,Albertus:2009ww,Faessler:2009xn,Onishchenko:2000wf,Kiselev:2001fw,White:1991hz,SanchisLozano:1994vh,Hernandez:2007qv,
Guo:1998yj,Ebert:2004ck,Wang:2017mqp}. However, only a few
discussions on the decay behavior exist in
literature~\cite{Hackman:1977am,Branz:2010pq,Bernotas:2013eia,Li:2017,Xiao:2017udy,Lu:2017meb}.
In our previous work~\cite{Xiao:2017udy}, we first systematically
investigated the both strong and radiative transitions of the
low-lying $1P$-wave doubly heavy baryons with chiral and constituent
quark model. In this work, we shall perform a systematic analysis of
the two-body Okubo-Zweig-Iizuka (OZI) allowed strong decays of the
$1P$ and $2D$ doubly charmed states with the quark pair
creation(QPC) model, which may provide more information of their
inner structures. The quark model classification, predicted
masses~\cite{Ebert:2002ig}, and OZI allowed decay
modes~\cite{Zhong:2007gp} are summarized in Table~\ref{mass}.

For the low-lying $1P$ and $2D$ doubly charmed baryons, their masses
are large enough to allow the decay channels containing a
heavy-light flavor meson. Thus, it is suitable to apply the QPC
strong decay model. Meanwhile, for further understanding the strong
decays of the doubly charmed baryons, it is necessary to make a
comparison of the theoretical predictions with QPC model to the
results with the chiral quark model~\cite{Xiao:2017udy}.

The QPC strong decay model as a phenomenological method has been
employed successfully in the description of the hadronic decays of
the
mesons~\cite{Godfrey:2015dva,Godfrey:2004ya,Godfrey:2015dia,Godfrey:2016nwn}
and singly charmed
baryons~\cite{Chen:2007xf,Chen:2017gnu,Zhao:2016qmh,Ye:2017dra,Ye:2017yvl}.
Systematical study of the low-lying $1P$ and $2D$ doubly charmed
states with the QPC model has not been performed yet. In the
framework of the QPC model, we find that (i) our results of the
decay patterns of the $1P$ states are highly comparable with those
in our previous work~\cite{Xiao:2017udy}; (ii) the $2D_{\rho\rho}$
states mainly decay via emitting a heavy-light meson if their masses
are above the threshold of $\Lambda_cD$ or $\Xi_cD$, respectively;
(iii) although the $2D_{\lambda\lambda}$ states may be broad states
with a width of $\Gamma>100$ MeV, they still have the opportunity to
be discovered via their main decay channels in future experiments.

This paper is structured as follows. In Sec. II we give a brief
review of the QPC model. We present our numerical results and
discussions in Sec. III and summarize our results in Sec. IV.

\begin{table*}[htpb]
\caption{\label{mass} Masses and possible two body strong decay
channels of the $1P$ and $2D$ doubly charmed baryons ( denoted by
$|N^{2S+1}L_{\sigma}J^P\rangle$), where
$|N^{2S+1}L_{\sigma}J^P\rangle$=$\sum_{L_z+S_z=J_z}\langle
LL_z,SS_z|JJ_z\rangle^N\Psi^{\sigma}_{LL_z}\chi_{S_z}\phi$
~\cite{Zhong:2007gp}. The masses (MeV) are taken from the
relativistic quark model~\cite{Ebert:2002ig}. }
\begin{tabular}{ccccccccccc}\hline\hline
State  & & \multicolumn{2}{c}{$\Xi_{cc}$}&
&\multicolumn{2}{c}{$\Omega_{cc}$} \\ \cline{3-4}\cline{6-7}
$N^{2S+1}L_{\sigma}J^P$  ~~& Wave function ~~&Mass~\cite{Ebert:2002ig}~~~&Strong decay channel  & &Mass~\cite{Ebert:2002ig}~~~&Strong decay channel   \\
 $|0^2S\frac{1}{2}^+\rangle$  &$^{0}\Psi^S_{00}\chi^\lambda_{S_z}\phi$   &3620~~~  & $\cdot\cdot\cdot$          & &3778 ~~~&$\cdot\cdot\cdot$ \\
 $|0^4S\frac{3}{2}^+\rangle$  &$^{0}\Psi^S_{00}\chi^s_{S_z}\phi$         &3727~~~  &$\cdot\cdot\cdot$   &&3872~~~&$\cdot\cdot\cdot$              \\
 $|1^2P_{\rho}\frac{1}{2}^-\rangle$&$^{1}\Psi^{\rho}_{1L_z} \chi^{\rho}_{S_z}\phi $ & 3838 ~~~&$\cdot\cdot\cdot$    &&4002~~~& $\cdot\cdot\cdot$            \\
 $|1^2P_{\rho}\frac{3}{2}^-\rangle$&         & 3959 ~~~&$\cdot\cdot\cdot$        &&4102~~~&$\cdot\cdot\cdot$     \\
 $|1^2P_{\lambda}\frac{1}{2}^-\rangle$ &$^{1}\Psi^{\lambda}_{1L_z} \chi^{\lambda}_{S_z}\phi$ & 4136 ~~~&$\Xi^{(*)}_{cc}\pi$   &&4271~~~&$\Xi^{(*)}_{cc}K$\\
 $|1^2P_{\lambda}\frac{3}{2}^-\rangle$  &    & 4196~~~&$\Xi^{(*)}_{cc}\pi$    &&4325~~~&$\Xi^{(*)}_{cc}K$         \\
 $|1^4P_{\lambda}\frac{1}{2}^-\rangle$  &$^{1}\Psi^{\lambda}_{1L_z} \chi^{s}_{S_z}\phi$  &4053~~~&$\Xi^{(*)}_{cc}\pi$    &&4208~~~&$\Xi_{cc}K$ \\
 $|1^4P_{\lambda}\frac{3}{2}^-\rangle$ &   & 4101~~~&$\Xi^{(*)}_{cc}\pi$   &&4252~~~&$\Xi^{(*)}_{cc}K$ \\
 $|1^4P_{\lambda}\frac{5}{2}^-\rangle$ &   & 4155~~~&$\Xi^{(*)}_{cc}\pi$   &&4303~~~&$\Xi^{(*)}_{cc}K$ \\
 $|2^2D_{\rho\rho}\frac{3}{2}^+\rangle$&$^{2}\Psi^{\rho\rho}_{2L_z} \chi^{\lambda}_{S_z}\phi$&~~~ &$\Lambda_cD$  && ~~~& $\Xi_cD$  \\
 $|2^2D_{\rho\rho}\frac{5}{2}^+\rangle$&   &     ~~~ &$\Lambda_cD$, $\Sigma_c^{(*)}D$    && ~~~& $\Xi_cD$,~$\Xi'^{(*)}_cD$         \\
 $|2^4D_{\rho\rho}\frac{1}{2}^+\rangle$&$^{2}\Psi^{\rho\rho}_{2L_z} \chi^{s}_{S_z}\phi$   & ~~~ &$\Lambda_cD$    && ~~~& $\Xi_cD$         \\
 $|2^4D_{\rho\rho}\frac{3}{2}^+\rangle$&   &     ~~~ &$\Lambda_cD$,~$\Sigma_cD$    && ~~~& $\Xi_cD$,~$\Xi'_cD$         \\
 $|2^4D_{\rho\rho}\frac{5}{2}^+\rangle$&   &     ~~~ &$\Lambda_cD$,~$\Sigma_c^{(*)}D$    && ~~~& $\Xi_cD$,~$\Xi'^*_cD$         \\
 $|2^4D_{\rho\rho}\frac{7}{2}^+\rangle$&   &  ~~~ &$\Lambda_cD$,~$\Sigma^{(*)}_cD$,~$\Xi_cD_s$,~$\Xi'_cD_s$    && ~~~& $\Xi_cD$,~$\Xi'^{(*)}_cD$,~$\Omega^{(*)}_cD_s$ \\
 $|2^2D_{\lambda\lambda}\frac{3}{2}^+\rangle$&$^{2}\Psi^{\lambda\lambda}_{2L_z} \chi^{\lambda}_{S_z}\phi$&~~~ &$\Xi^{(*)}_{cc}\pi$,~$\Omega^{(*)}_{cc}K$,~$\Lambda_cD$,~$\Sigma^{(*)}_cD$,~$\Xi_cD_s$,~$\Xi'^{(*)}_cD_s$  && ~~~& $\Xi^{(*)}_{cc}K$,~$\Omega^{(*)}_{cc}\eta$,~$\Xi_cD$~$\Xi'^{(*)}_cD$,~$\Omega^{(*)}_cD_s$,~$\Omega^{(*)}_{cc}\eta'$  \\
 $|2^2D_{\lambda\lambda}\frac{5}{2}^+\rangle$& &~~~ &$\Xi^{(*)}_{cc}\pi$,~$\Omega^{(*)}_{cc}K$,~$\Lambda_cD$,~$\Sigma^{(*)}_cD$,~$\Xi_cD_s$,~$\Xi'^{(*)}_cD_s$  && ~~~& $\Xi^{(*)}_{cc}K$,~$\Omega^{(*)}_{cc}\eta$,~$\Xi_cD$,~$\Xi'^{(*)}_cD$,~$\Omega^{(*)}_cD_s$,~$\Omega^{(*)}_{cc}\eta'$  \\
 $|2^4D_{\lambda\lambda}\frac{1}{2}^+\rangle$&$^{2}\Psi^{\lambda\lambda}_{2L_z} \chi^{s}_{S_z}\phi$&~~~ &$\Xi^{(*)}_{cc}\pi$,~$\Omega^{(*)}_{cc}K$,~$\Lambda_cD$,~$\Sigma^{(*)}_cD$,~$\Xi_cD_s$,~$\Xi'_cD_s$  && ~~~& $\Xi^{(*)}_{cc}K$,~$\Omega^{(*)}_{cc}\eta$,~$\Xi_cD$,~$\Xi'^{(*)}_cD$,~$\Omega_cD_s$  \\
 $|2^4D_{\lambda\lambda}\frac{3}{2}^+\rangle$&&~~~ &$\Xi^{(*)}_{cc}\pi$,~$\Omega^{(*)}_{cc}K$,~$\Lambda_cD$,~$\Sigma^{(*)}_cD$,~$\Xi_cD_s$,~$\Xi'^{(*)}_cD_s$  && ~~~& $\Xi^{(*)}_{cc}K$,~$\Omega^{(*)}_{cc}\eta$,~$\Omega_{cc}\eta'$,~$\Xi_cD$,~$\Xi'^{(*)}_cD$,~$\Omega^{(*)}_cD_s$  \\
 $|2^4D_{\lambda\lambda}\frac{5}{2}^+\rangle$&&~~~ &$\Xi^{(*)}_{cc}\pi$,~$\Omega^{(*)}_{cc}K$,~$\Lambda_cD$,~$\Sigma^{(*)}_cD$,~$\Xi_cD_s$,~$\Xi'^{(*)}_cD_s$  && ~~~& $\Xi^{(*)}_{cc}K$,~$\Omega^{(*)}_{cc}\eta$,~$\Omega_{cc}\eta'$,~$\Xi_cD$,~$\Xi'^{(*)}_cD$,~$\Omega^{(*)}_cD_s$  \\
 $|2^4D_{\lambda\lambda}\frac{7}{2}^+\rangle$&&~~~ &$\Xi^{(*)}_{cc}\pi$,~$\Omega^{(*)}_{cc}K$,~$\Lambda_cD$,~$\Sigma^{(*)}_cD$,~$\Xi_cD_s$,~$\Xi'^{(*)}_cD_s$  && ~~~& $\Xi^{(*)}_{cc}K$,~$\Omega^{(*)}_{cc}\eta$,~$\Omega^{(*)}_{cc}\eta'$,~$\Xi_cD$,~$\Xi'^{(*)}_cD$,~$\Omega^{(*)}_cD_s$  \\
\hline\hline
\end{tabular}
\end{table*}

\section{$^3P_0$ model}\label{model}

The QPC model was first proposed by Micu~\cite{Micu:1968mk}, Carlitz
and Kislinger~\cite{Carlitz:1970xb}, and further developed by the
Orsay
group~\cite{LeYaouanc:1972vsx,LeYaouanc:1988fx,LeYaouanc:1977fsz}.
For the OZI-allowed strong decays of hadrons, this model assumes
that a pair of quark $q\bar{q}$ is created from the vacuum and then
regroups with the quarks from the initial hadron to produce two
outing hadrons. The created quark pair $q\bar{q}$ shall carry the
quantum number of $0^{++}$ and be in a $^3P_0$ state. Thus the QPC
model is also known as the $^3P_0$ model. This model has been
extensively employed to study the OZI-allowed strong transitions of
hadron systems. Here, we adopt this model to study the strong decays
of the $ccq$ system.

\begin{figure}[htpb]
\centering \epsfxsize=7.5 cm \epsfbox{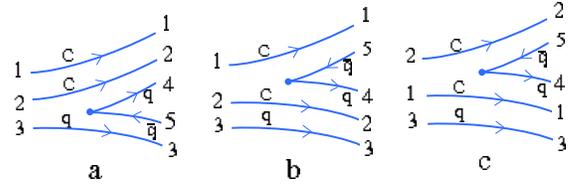} \caption{Doubly
charmed baryons decay process in the $^3P_0$ model.}\label{qkp}
\end{figure}

According to the quark rearrangement process, any of the three
quarks in the initial baryon can go into the final meson. Thus three
possible decay processes are take into account as shown in
Fig.~\ref{qkp}. Now, we take the Fig.~\ref{qkp}(a) decay process
$A$(the initial baryon)$\rightarrow$ $B$(the final baryon)+$C$(the
final meson) as an example to show how to calculate the decay width.
In the nonrelativistic limit, the transition operator under the
$^3P_0$ model is given by
\begin{eqnarray}
T&=&-3\gamma \sum_m\langle1m;1-m|00\rangle\int
d^3\mathbf{p}_4d^3\mathbf{p}_5\delta^3(\mathbf{p}_4+\mathbf{p}_5)\\
\nonumber &&\times\omega^{45}_0\varphi^{45}_0\chi^{45}_{1,-m}
\mathcal{Y}_1^m(\frac{\mathbf{p}_4-\mathbf{p}_5}{2})a^{\dagger}_{4i}b^{\dagger}_{5j},
\end{eqnarray}
where $\mathbf{p}_i$ ($i$=4, 5) represents the three-vector momentum
of the $i$th quark in the created quark pair.
$\omega^{45}_0=\delta_{ij}$ and
$\varphi^{45}_0=(u\bar{u}+d\bar{d}+s\bar{s})/\sqrt{3}$ stand for the
color singlet and flavor function, respectively. The solid harmonic
polynomial
$\mathcal{Y}_1^m(\mathbf{p})\equiv|\mathbf{p}|Y^m_1(\theta_p,\phi_p)$
corresponds to the momentum-space distribution, and
$\chi^{45}_{1,-m}$ is the spin triplet state for the created quark
pair. The creation operator $a^{\dagger}_{4i}b^{\dagger}_{5j}$
denotes the quark pair-creation in the vacuum. The pair-creation
strength $\gamma$ is a dimensionless parameter, which is usually
fixed by fitting the well measured partial decay widths.

According to the definition of the mock state~\cite{Hayne:1981zy},
the spacial wave functions of the baryon and meson read,
respectively,
\begin{eqnarray}
|A(N_A~^{2S_A+1}L_A{J_AM_{J_A}})(\mathbf{p}_A)\rangle=~~~~~~~~~~~~~~~~~~~~~~~~~~~~~~~~~~~~~~\nonumber \\
\sqrt{2E_A}\varphi^{123}_A\omega^{123}_A\sum_{M_{L_A},M_{S_A}}\langle L_AM_{L_A};S_AM_{S_A}|J_AM_{J_A}\rangle\nonumber \\
\times\int d^3\mathbf{p}_1d^3\mathbf{p}_2d^3\mathbf{p}_3\delta^3(\mathbf{p}_1+\mathbf{p}_2+\mathbf{p}_3-\mathbf{p}_A)\nonumber \\
\times\Psi_{N_AL_AM_{L_A}(\mathbf{p}_1,\mathbf{p}_2,\mathbf{p}_3)}\chi^{123}_{S_AM_{S_A}}
|q_1(\mathbf{p}_1)q_2(\mathbf{p}_2)q_3(\mathbf{p}_3)\rangle,
\end{eqnarray}
\begin{eqnarray}
|C(N_C~^{2S_C+1}L_C{J_CM_{J_C}})(\mathbf{p}_C)\rangle=~~~~~~~~~~~~~~~~~~~~~~~~~~~~~~~~~~~~~~\nonumber \\
\sqrt{2E_C}\varphi^{ab}_C\omega^{ab}_C\sum_{M_{L_C},M_{S_C}}\langle L_CM_{L_C};S_CM_{S_C}|J_CM_{J_C}\rangle\nonumber \\
\times\int d^3\mathbf{p}_ad^3\mathbf{p}_b\delta^3(\mathbf{p}_a+\mathbf{p}_b-\mathbf{p}_C)\nonumber \\
\times\Psi_{N_CL_CM_{L_C}(\mathbf{p}_a,\mathbf{p}_b)}\chi^{ab}_{S_CM_{S_C}}|q_a(\mathbf{p}_a)q_b(\mathbf{p}_b)\rangle.
\end{eqnarray}
The $\mathbf{p}_i(i=1,2,3~\text{and}~a,b)$ denotes the momentum of
quarks in hadron $A$ and $C$. $P_A(P_C)$ are the momentum of the
hadron $A(C)$. The $^3P_0$ model gives a good description of the
decay properties of many observed mesons with the simple harmonic
oscillator space-wave functions, which are adopted to describe the
spatial wave function of both baryons and mesons in the present
work. The spatial wave function of a baryon without the radial
excitation is
\begin{eqnarray}
\psi^0_{lm}(\mathbf{p})=(-i)^l\Bigg[\frac{2^{l+2}}{\sqrt{\pi}(2l+1)!!}\Bigg]^{\frac{1}{2}}\Bigg(\frac{1}{\alpha}\Bigg)^{l+\frac{3}{2}}
\text{exp}\Bigg(-\frac{\mathbf{p}^2}{2\alpha^2}\Bigg)\mathcal{Y}_l^m(\mathbf{p}).
\end{eqnarray}
The ground state spatial wave function of a meson is
\begin{eqnarray}
\psi_{0,0}=\Bigg(\frac{R^2}{\pi}\Bigg)^{\frac{3}{4}}\text{exp}\Bigg(-\frac{R^2\mathbf{p}_{ab}^2}{2}\Bigg),
\end{eqnarray}
where the $\mathbf{p}_{ab}$ stands for the relative momentum between
the quark and antiquark in the meson. Then, we can obtain the
partial decay amplitude in the center of mass frame,
\begin{eqnarray}
\mathcal{M}^{M_{J_A}M_{J_B}M_{J_C}}(A\rightarrow B+C)&=&\gamma\sqrt{8E_AE_BE_C}\nonumber\\
&&\prod_{A,B,C}\langle\chi^{124}_{S_BM_{S_B}}\chi^{35}_{S_CM_{S_C}}|\chi^{123}_{S_AM_{S_A}}\chi^{45}_{1-m}\rangle
\nonumber\\
&&\langle\varphi^{124}_B\varphi^{35}_C|\varphi^{123}_A\varphi^{45}_0\rangle
I^{M_{L_A},m}_{M_{L_B},M_{L_C}}(\mathbf{p}).
\end{eqnarray}
Here, $I^{M_{L_A},m}_{M_{L_B},M_{L_C}}(\mathbf{p})$ stands the
spatial integral and more detailed information is presented in the
Appendix A and B. The $\prod_{A,B,C}$ denotes the Clebsch-Gorden
coefficients for the quark pair, initial and final hadrons, which
come from the couplings among the orbital, spin, and total angular
momentum. Its expression reads
\begin{eqnarray}
\sum\langle L_BM_{L_B};S_BM_{S_B}|J_BM_{J_B}\rangle\langle L_CM_{L_C};S_CM_{S_C}|J_CM_{J_C}\rangle\nonumber \\
\times\langle L_AM_{L_A};S_AM_{S_A}|J_AM_{J_A}\rangle\langle
1m;1-m|00\rangle
\end{eqnarray}

Finally, the decay width $\Gamma[A\rightarrow BC]$ reads
\begin{eqnarray}
\Gamma[A\rightarrow
BC]=\pi^2\frac{|\mathbf{p}|}{M_A^2}\frac{1}{2J_A+1}\sum_{M_{J_A},M_{J_B},M_{J_C}}|\mathcal{M}^{M_{J_A}M_{J_B}M_{J_C}}|^2.
\end{eqnarray}
In the equation, $\mathbf{p}$ is the momentum of the daughter baryon
in the center of mass frame of the parent baryon $A$
\begin{eqnarray}
|\mathbf{p}|=\frac{\sqrt{[M_A^2-(M_B-M_C)^2][M_A^2-(M_B+M_C)^2]}}{2M_A}
\end{eqnarray}

\begin{table}[htpb]
\caption{\label{finalstates} Masses (MeV) of the baryons and mesons
in the decays~\cite{Aaij:2017ueg,Olive:2016xmw,Ebert:2002ig}. }
\begin{tabular}{ccccccccccc}\hline\hline
State~~~~~~&Mass~~~~~~&State~~~~~~&Mass~~~~~~&State~~~~~~&Mass \\
$\Xi^{++(+)}_{cc}$~~~~~~&3621.00~~~~~~&$\Sigma^{++}_c$~~~~~~&2453.97~~~~~~&$\pi^0$ &134.977 \\
$\Xi^{*++(+)}_{cc}$~~~~~~&3727.00~~~~~~&$\Sigma^{*++}_c$~~~~~~&2518.41~~~~~~&$\pi^+$ &139.570 \\
$\Omega^+_{cc}$~~~~~~&3778.00~~~~~~&$\Sigma^+_c$~~~~~~&2452.90~~~~~~&$K^{\pm}$ &493.677 \\
$\Omega^{*+}_{cc}$~~~~~~&3872.00~~~~~~&$\Sigma^{*+}_c$~~~~~~&2517.50~~~~~~&$\eta$ &547.862 \\
$\Lambda^{+}_{c}$~~~~~~&2286.46~~~~~~&$\Xi^{+}_c$~~~~~~&2467.93~~~~~~&$\eta'$ &957.780 \\
$\Omega^{0}_{c}$~~~~~~&2695.20~~~~~~&$\Xi'^{+}_c$~~~~~~&2575.70~~~~~~&$D^0$ &1864.83 \\
$\Omega^{*0}_{c}$~~~~~~&2765.90~~~~~~&$\Xi'^{*+}_c$~~~~~~&2645.90~~~~~~&$D^+$ &1869.58 \\
                 ~~~~~~&       ~~~~~~&             ~~~~~~&       ~~~~~~&$D^+_s$ &1968.27 \\
\hline\hline
\end{tabular}
\end{table}

In the present calculation, we adopt $m_u=m_d=220$ MeV, $m_s=419$
MeV, and $m_c=1628$ MeV~\cite{Godfrey:2015dva} for the constituent
quark masses. The masses of the baryons and mesons involved in our
calculations, listed in Table~\ref{finalstates}, are from the
Particle Data Group~\cite{Olive:2016xmw} except for the doubly
charmed baryons, which is from Ref.~\cite{Ebert:2002ig}. The value
of the harmonic oscillator strength $R$ is 2.5 $\text{GeV}^{-1}$,
for all light flavor mesons while it is $R=1.67 \text{GeV}^{-1}$ for
the $D$ meson and $R=1.54 \text{GeV}^{-1}$ for the $D_s$
meson~\cite{Godfrey:2015dva}. The parameter $\alpha_{\rho}$ of the
$\rho$-mode excitation between the two charm quarks is taken as
$\alpha_{\rho}=0.66$ GeV~\cite{Xiao:2017udy}, while $\alpha_{\rho}$
between the two light quarks is taken as $\alpha_{\rho}=0.4$ GeV.
Another harmonic oscillator parameter $\alpha_{\lambda}$ is obtained
with the relation:
\begin{equation}
\alpha_{\lambda}=\Bigg(\frac{3m_3}{2m_1+m_3}\Bigg)^{1/4}\alpha_{\rho}.
\end{equation}
For the strength of the quark pair creation from the vacuum, we take
the same value as in Ref.~\cite{Godfrey:2015dva}, $\gamma=6.95$. For
the strange quark pair $s\bar{s}$ creation, we use
$\gamma_{s\bar{s}}=\gamma/\sqrt{3}$~\cite{LeYaouanc:1977fsz}.

\begin{table*}[htpb]
\caption{\label{Pwave} The comparison of the partial decay widths of
the $1P_{\lambda}$ states from the QPC model and the chiral quark
model ~\cite{Xiao:2017udy}. $\Gamma_{\text{total}}$ stands for the
total decay width and $\mathcal{B}$ represent the ratio of the
branching fractions $\Gamma[\Xi_{cc}\pi/K]/\Gamma[\Xi^*_{cc}\pi/K]$.
The unit is MeV.}
\begin{tabular}{cccccccccccccccccc}\hline\hline
~~~~~~&~~~~~~&\multicolumn{2}{c}{$\Gamma[\Xi_{cc}\pi]$}&~~~~~~&\multicolumn{2}{c}{$\Gamma[\Xi^*_{cc}\pi]$}&~~~~~~&\multicolumn{2}{c}{Total}
&~~~~~~&\multicolumn{2}{c}{$\mathcal{B}$}\\
\cline{3-4}\cline{6-7}\cline{9-10}\cline{12-13}
State~~~~~~&Mass~~~~~~&This work~~~~~~&Ref~\cite{Xiao:2017udy}&~~~~~~&This work~~~~~~&Ref~\cite{Xiao:2017udy}&~~~~~~&This work~~~~~~&Ref~\cite{Xiao:2017udy}&~~~~~~&This work~~~~~~&Ref~\cite{Xiao:2017udy}\\
$|\Xi_{cc}~^2P_{\lambda}\frac{1}{2}^-\rangle$~~~~~~&4136~~~~~~&21.9~~~~~~&15.6&~~~~~~&18.6~~~~~~&33.9&~~~~~~&40.5~~~~~~&49.5&~~~~~~&1.18~~~~~~&0.46\\
$|\Xi_{cc}~^2P_{\lambda}\frac{3}{2}^-\rangle$~~~~~~&4196~~~~~~&13.7~~~~~~&21.6&~~~~~~&117~~~~~~&101&~~~~~~&131~~~~~~&123&~~~~~~&0.18~~~~~~&0.21\\
$|\Xi_{cc}~^4P_{\lambda}\frac{1}{2}^-\rangle$~~~~~~&4053~~~~~~&200~~~~~~&133&~~~~~~&0.60~~~~~~&1.22&~~~~~~&201~~~~~~&134&~~~~~~&333~~~~~~&110\\
$|\Xi_{cc}~^4P_{\lambda}\frac{3}{2}^-\rangle$~~~~~~&4101~~~~~~&4.43~~~~~~&7.63&~~~~~~&127~~~~~~&84.6&~~~~~~&131~~~~~~&92.2&~~~~~~&0.03~~~~~~&0.09\\
$|\Xi_{cc}~^4P_{\lambda}\frac{5}{2}^-\rangle$~~~~~~&4155~~~~~~&45.9~~~~~~&75.3&~~~~~~&12.6~~~~~~&22.8&~~~~~~&58.5~~~~~~&98.1&~~~~~~&3.64~~~~~~&3.30\\
\hline\hline
~~~~~~&~~~~~~&\multicolumn{2}{c}{$\Gamma[\Xi_{cc}K]$}&~~~~~~&\multicolumn{2}{c}{$\Gamma[\Xi^*_{cc}K]$}&~~~~~~&\multicolumn{2}{c}{Total}
&~~~~~~&\multicolumn{2}{c}{$\mathcal{B}$}\\
\cline{3-4}\cline{6-7}\cline{9-10}\cline{12-13}
State~~~~~~&Mass~~~~~~&This work~~~~~~&Ref~\cite{Xiao:2017udy}&~~~~~~&This work~~~~~~&Ref~\cite{Xiao:2017udy}&~~~~~~&This work~~~~~~&Ref~\cite{Xiao:2017udy}&~~~~~~&This work~~~~~~&Ref~\cite{Xiao:2017udy}\\
$|\Omega_{cc}~^2P_{\lambda}\frac{1}{2}^-\rangle$~~~~~~&4271~~~~~~&49.3~~~~~~&33.1&~~~~~~&1.53~~~~~~&2.36&~~~~~~&50.8~~~~~~&35.5&~~~~~~&32.2~~~~~~&14.0\\
$|\Omega_{cc}~^2P_{\lambda}\frac{3}{2}^-\rangle$~~~~~~&4325~~~~~~&8.50~~~~~~&11.4&~~~~~~&199~~~~~~&174&~~~~~~&208~~~~~~&185&~~~~~~&0.04~~~~~~&0.06\\
$|\Omega_{cc}~^4P_{\lambda}\frac{1}{2}^-\rangle$~~~~~~&4208~~~~~~&378~~~~~~&323&~~~~~~&$\cdot\cdot\cdot$~~~~~~&$\cdot\cdot\cdot$&~~~~~~&378~~~~~~&323
&~~~~~~&$\cdot\cdot\cdot$~~~~~~&$\cdot\cdot\cdot$\\
$|\Omega_{cc}~^4P_{\lambda}\frac{3}{2}^-\rangle$~~~~~~&4252~~~~~~&2.02~~~~~~&3.08&~~~~~~&154~~~~~~&137&~~~~~~&156~~~~~~&140&~~~~~~&0.01~~~~~~&0.02\\
$|\Omega_{cc}~^4P_{\lambda}\frac{5}{2}^-\rangle$~~~~~~&4303~~~~~~&29.1~~~~~~&41.5&~~~~~~&2.62~~~~~~&4.38&~~~~~~&31.7~~~~~~&45.9&~~~~~~&11.1~~~~~~&9.47\\
\hline\hline
\end{tabular}
\end{table*}

\section{Calculations and Results }\label{results}
For the $P$-wave doubly charmed states, the masses are adopted from
Ref.~\cite{Ebert:2002ig} (showed in Table~\ref{mass}) due to a good
agreement with the mass of the lowest doubly charmed baryon
$\Xi^{++}_{cc}(3621)$ observed by the LHCb collaboration. However,
there is no prediction for the masses of the $D$-wave states. So the
masses of the $D$-wave baryons are varied in a rough range when
their decay properties are investigated.

\subsection{The $P$-wave doubly charmed states}
Within the quark model, there are two $1P_{\rho}$ doubly heavy
baryons with $J^P=\frac{1}{2}^-$ and $J^P=\frac{3}{2}^-$,
respectively. Their masses are above the threshold of $\Xi_{cc}\pi$
or $\Xi_{cc}K$. However, the OZI-allowed two body strong decays are
forbidden since the spatial wave functions for the $1P$ and $0S$
states are adopted with the simple harmonic oscillator wave
functions which are orthogonal. In this work, we focus on the strong
decays of the $1P_{\lambda}$ states.

We analyze the decay properties of the $1P_{\lambda}$ states in the
$\Xi_{cc}$ and $\Omega_{cc}$ family, and collect their partial
strong decay widths in Table~\ref{Pwave}. In the $\Xi_{cc}$ family,
the total decay width of
$|\Xi_{cc}~^2P_{\lambda}\frac{1}{2}^-\rangle$ is about $\Gamma\sim40$
MeV, which is compatible with the result in
Ref.~\cite{Xiao:2017udy}. The dominant decay modes are $\Xi_{cc}\pi$
and $\Xi^*_{cc}\pi$ with the partial decay ratio
\begin{equation}
\frac{\Gamma[|\Xi_{cc}~^2P_{\lambda}\frac{1}{2}^-\rangle\to\Xi_{cc}\pi]}{\Gamma[|\Xi_{cc}~^2P_{\lambda}\frac{1}
{2}^-\rangle\to\Xi^*_{cc}\pi]}\simeq1.18.
\end{equation}
This value is about 2.5 times of the ratio in
Ref.~\cite{Xiao:2017udy}.

The states of $|\Xi_{cc}~^2P_{\lambda}\frac{3}{2}^-\rangle$ and
$|\Xi_{cc}~^4P_{\lambda}\frac{3}{2}^-\rangle$ are most likely to be
the moderate states with a width of $\Gamma\sim130$ MeV, and the
$\Xi_{cc}^*\pi$ decay channel is their dominant decay mode. The
partial decay width of
$\Gamma[|\Xi_{cc}$$^2P_{\lambda}\frac{3}{2}^-\rangle\to
\Xi_{cc}\pi]$ is considerable. The partial decay width ratio is
\begin{eqnarray}
\frac{\Gamma[|\Xi_{cc}~^2P_{\lambda}3/2^-\rangle\rightarrow
\Xi_{cc}\pi]} {\Gamma[|\Xi_{cc}~^2P_{\lambda}3/2^-\rangle\rightarrow
\Xi^*_{cc}\pi]}\simeq 0.18.
\end{eqnarray}
This ratio may be a useful distinction between
$|\Xi_{cc}~^2P_{\lambda}\frac{3}{2}^-\rangle$ and
$|\Xi_{cc}~^4P_{\lambda}\frac{3}{2}^-\rangle$ in future experiments.
These results are in good agreement with the predictions in
Ref.~\cite{Xiao:2017udy}.

The state $|\Xi_{cc}~^4P_{\lambda}\frac{1}{2}^-\rangle$ has a broad
width of $\Gamma\simeq201$ MeV, and the $\Xi_{cc}\pi$ decay channel
almost saturates its total decay widths. This broad state may be
observed in $\Xi_{cc}\pi$ channel in future experiments.

\begin{figure}[htpb]
\centering \epsfxsize=8.0 cm \epsfbox{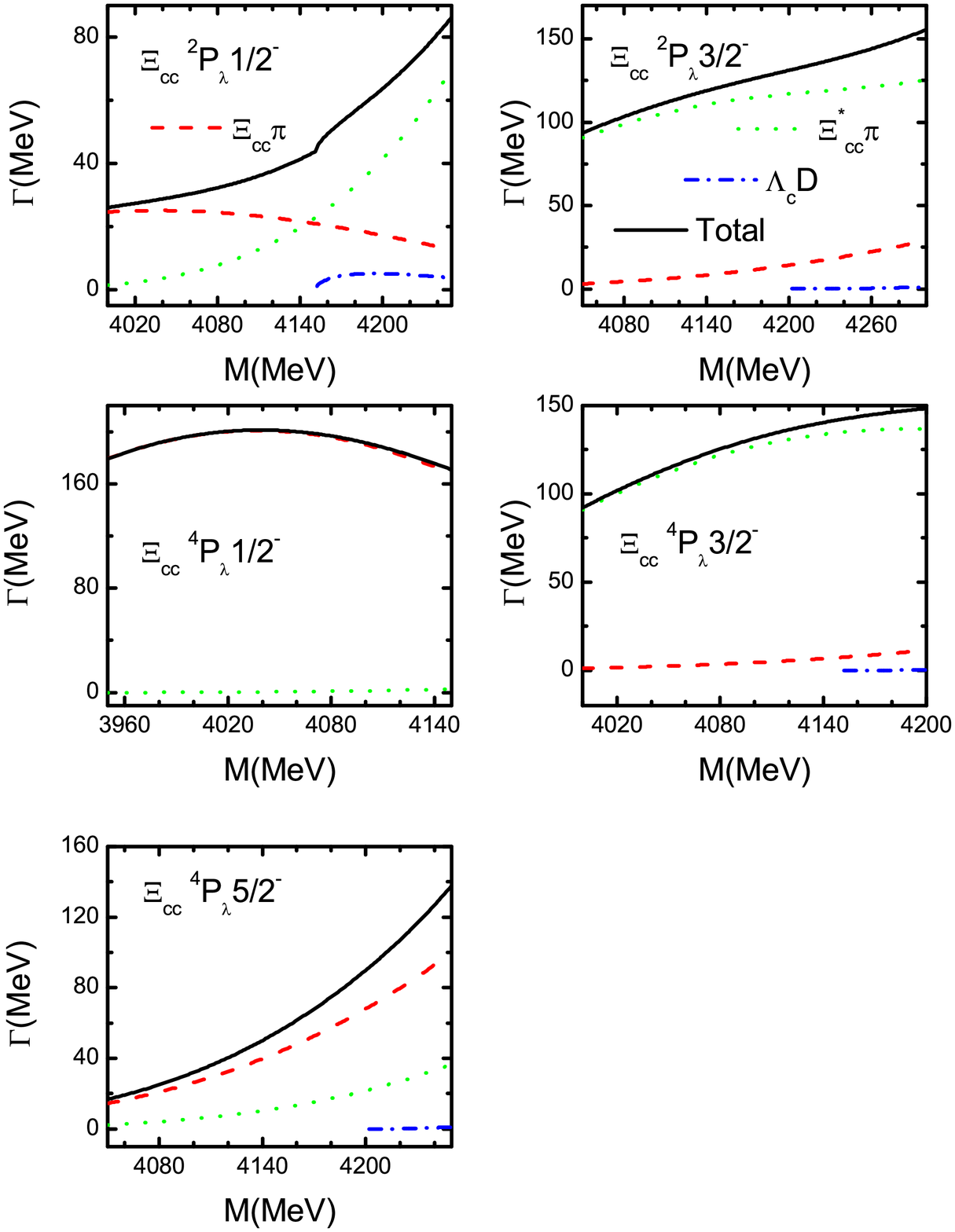} \caption{The
strong decay partial widths of the $1P_{\lambda}$-wave $\Xi_{cc}$
states as a function of the mass.}\label{XiP}
\end{figure}

\begin{figure}[htpb]
\centering \epsfxsize=8.0 cm \epsfbox{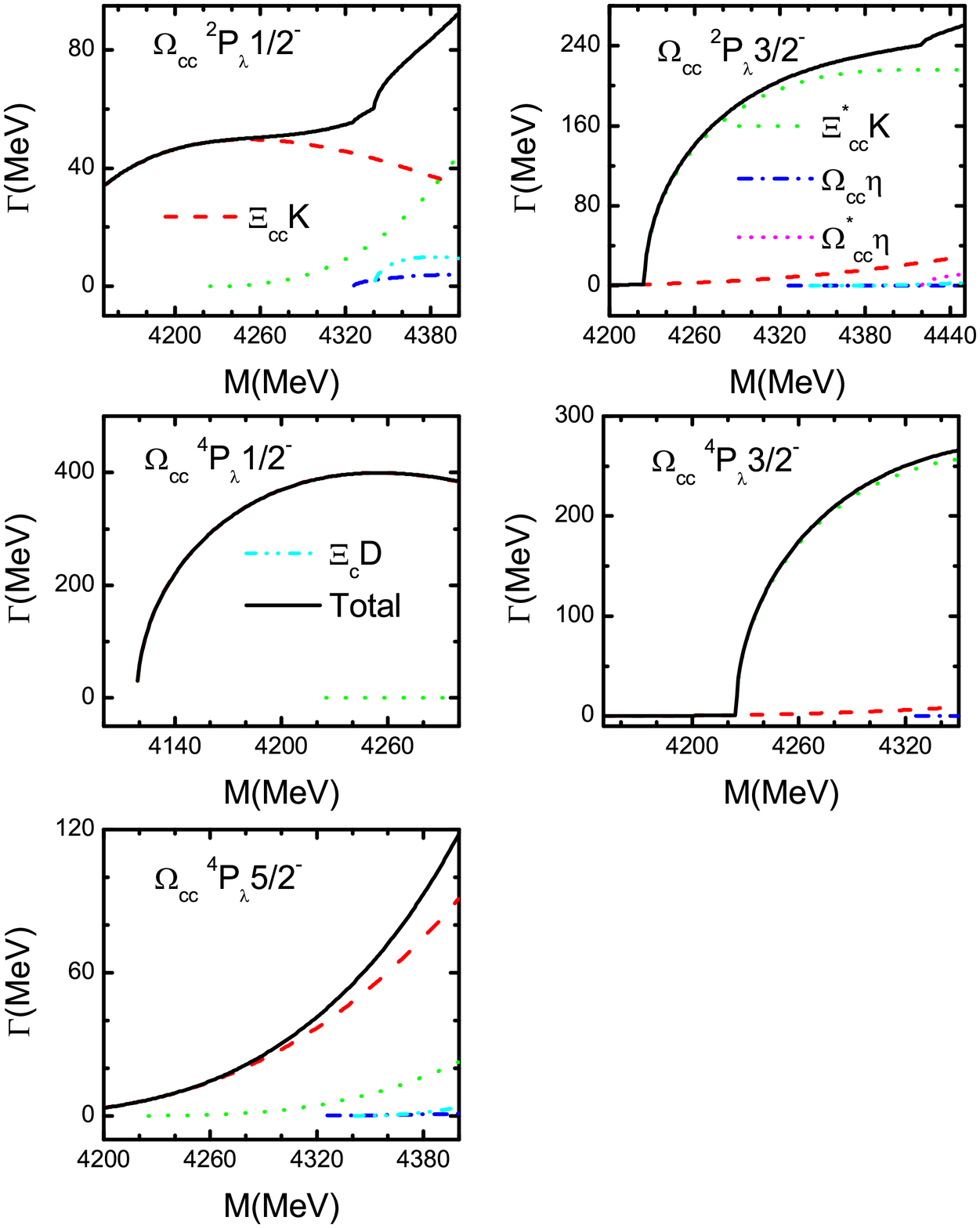} \caption{The
strong decay partial widths of the $1P_{\lambda}$-wave $\Omega_{cc}$
states as a function of the mass.}\label{OMGP}
\end{figure}

\begin{figure*}[htpb]
\centering \epsfxsize=12.0 cm \epsfbox{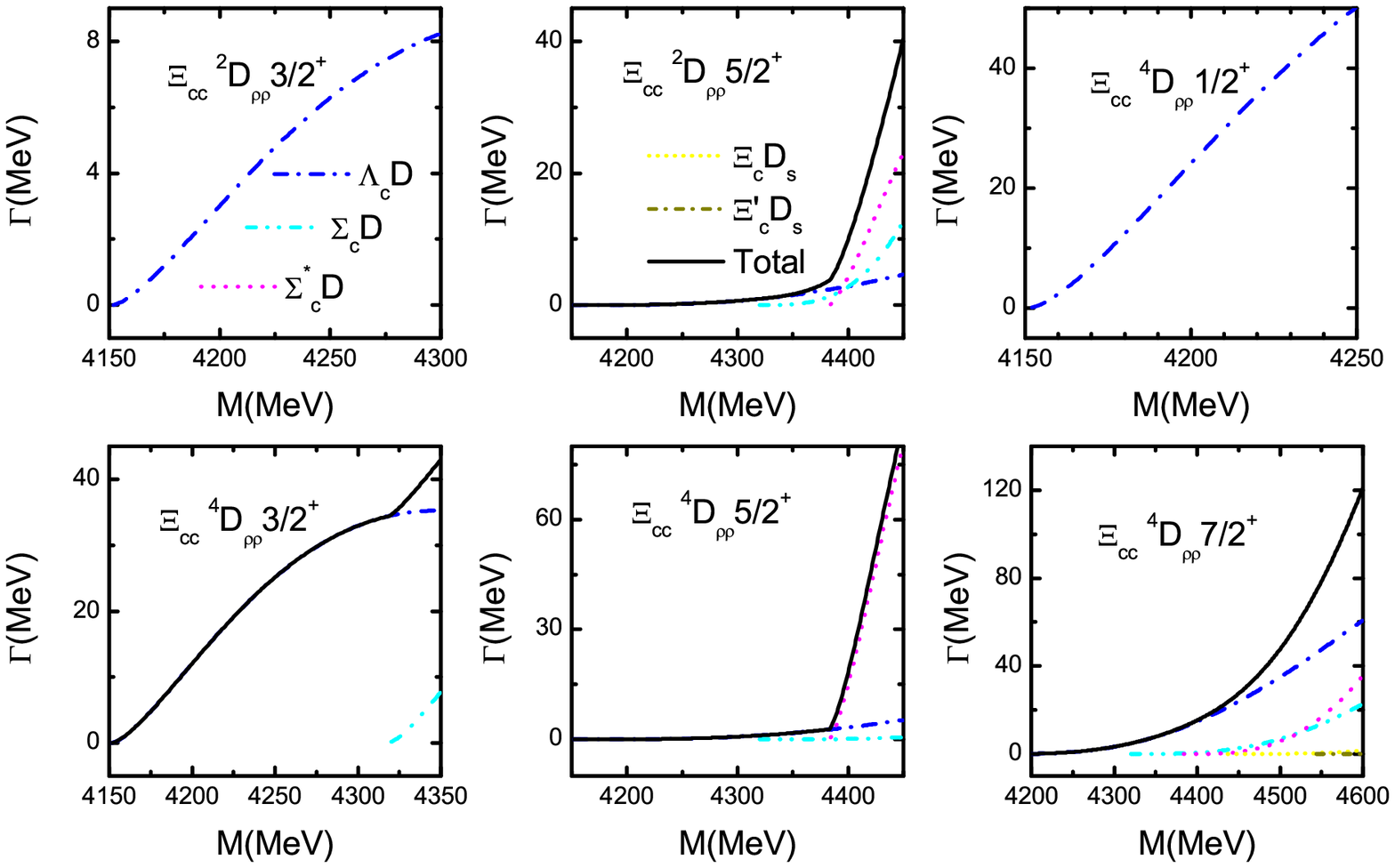} \caption{The
strong decay partial widths of the $2D_{\rho\rho}$-wave $\Xi_{cc}$
states as a function of the mass.}\label{XiDR}
\end{figure*}

From the Table~\ref{Pwave}, the state
$|\Xi_{cc}~^4P_{\lambda}\frac{5}{2}^-\rangle$ may be a narrow state
with a total decay width around $\Gamma\sim$60 MeV, which is about
one half of that in Ref.~\cite{Xiao:2017udy}. This state decays
mainly through the $\Xi_{cc}\pi$ channel. The predicted partial
width ratio between $\Xi_{cc}\pi$ and $\Xi_{cc}^*\pi$ is
\begin{equation}
\frac{\Gamma[|\Xi_{cc}~^4P_{\lambda}\frac{5}{2}^-\rangle\to\Xi_{cc}\pi]}{\Gamma[|\Xi_{cc}~^4P_{\lambda}\frac{5}
{2}^-\rangle\to\Xi^*_{cc}\pi]}\simeq3.64,
\end{equation}
which can be tested in future experiments.

In the $\Omega_{cc}$ family, the
$|\Omega_{cc}~^2P_{\lambda}\frac{1}{2}^-\rangle$ and
$|\Omega_{cc}~^4P_{\lambda}\frac{5}{2}^-\rangle$ might be two narrow
states with a total decay width of $\Gamma\sim$40 MeV, and their
strong decays are dominated by the $\Xi_{cc}K$ channel.

The decay width of the state
$|\Omega_{cc}~^4P_{\lambda}\frac{1}{2}^-\rangle$ is about
$\Gamma\sim$380 MeV. Meanwhile, its strong decays are governed by
the $\Xi_{cc}K$ channel. In this case, the
$|\Omega_{cc}~^4P_{\lambda}\frac{1}{2}^-\rangle$ might be too broad
to observed in experiments. However, for the states
$|\Omega_{cc}~^2P_{\lambda}\frac{3}{2}^-\rangle$ and
$|\Omega_{cc}~^4P_{\lambda}\frac{3}{2}^-\rangle$, if their masses lie
below the threshold of $\Xi^*_{cc}K$, they are likely to be two
fairly narrow states with the total decay widths of $\Gamma\sim9$
MeV and $\Gamma\sim2$ MeV, respectively. Otherwise, they shall have
a broad width of $\Gamma\sim$200 MeV, and mainly decay into
$\Xi^*_{cc}K$ channel.

Considering the mass uncertainties of the $1P_{\lambda}$ states, we
plot the strong decay width as a function of the mass in
Figs.~\ref{XiP}-~\ref{OMGP}. From the Figs.~\ref{XiP}-~\ref{OMGP},
the partial width of dominant decay channel for most of states are
sensitive to the mass. In addition, in the $\Xi_{cc}$ family, if the
$1P_{\lambda}$ states are above the threshold of $\Lambda_cD$, they
can decay via $\Lambda_cD$ with a partial width about several
MeV.

\begin{figure*}[htpb]
\centering \epsfxsize=12.0 cm \epsfbox{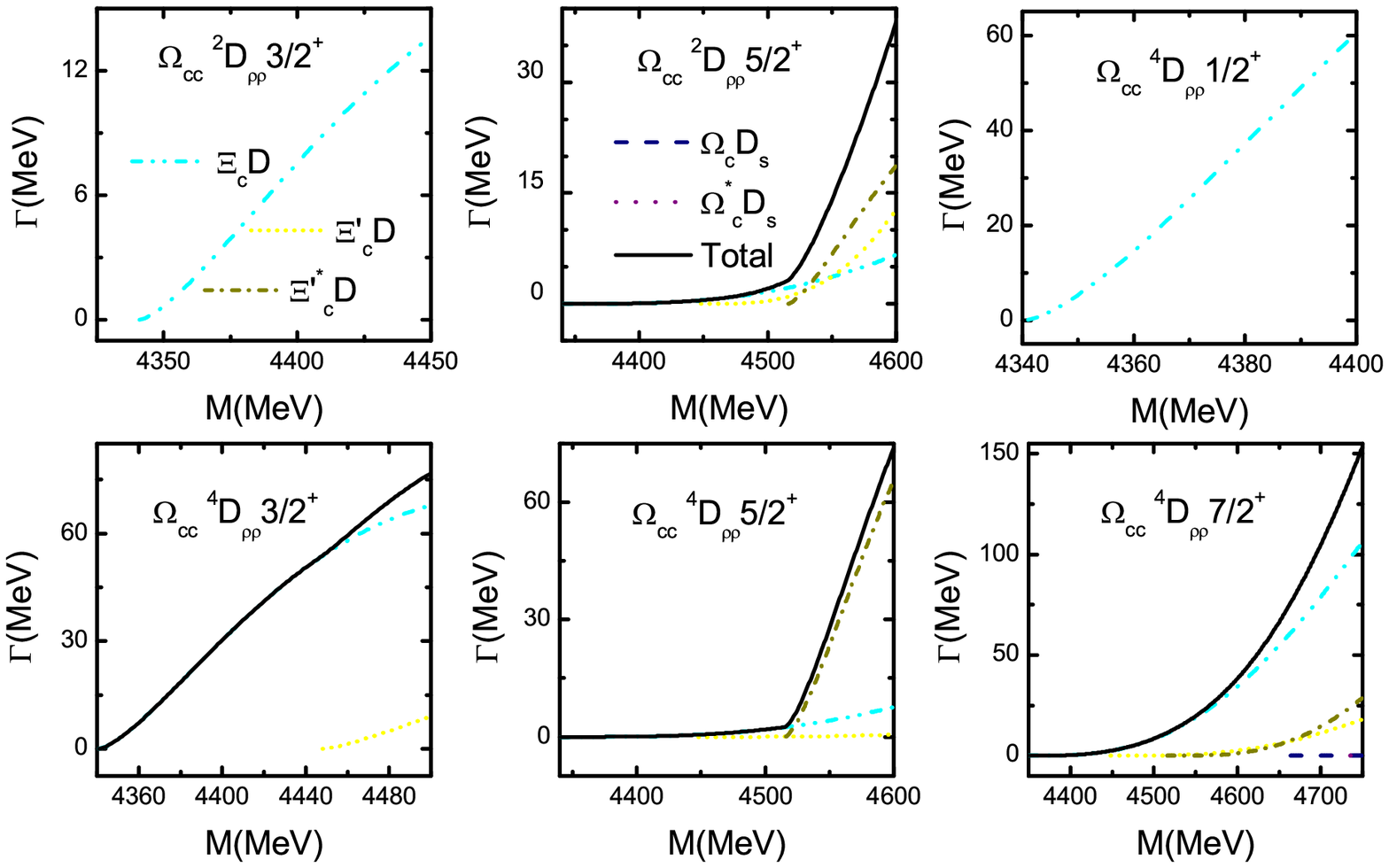} \caption{The
strong decay partial widths of the $2D_{\rho\rho}$-wave
$\Omega_{cc}$ states as a function of the mass.}\label{OMGDR}
\end{figure*}

\subsection{The $D$-wave doubly charmed states}
\subsubsection{$\rho$-mode excitations}

Since we adopt the simple harmonic oscillator spatial wave functions
in present work, the strong decays of $2D_{\rho\rho}$ doubly charmed
states via emitting a light-flavor meson are forbidden due to the
orthogonality of the spatial wave functions. So, we focus on the
decay processes via emitting a heavy-light flavor meson. Due to the
lack of the mass predictions for the $D$-wave doubly charmed states,
we investigate the strong decay properties as the functions of the
masses in a possible range.

First of all, we conduct systematic research on the strong decays of
$2D_{\rho\rho}$ states in the $\Xi_{cc}$ family in Fig.~\ref{XiDR}.
For the state $|\Xi_{cc}~^2D_{\rho\rho}\frac{3}{2}^+\rangle$, we put
the mass range between the $\Lambda_cD$ threshold ($M=4152$ MeV) and
$M=4300$ MeV. From Fig.~\ref{XiDR}, we can see that the state
$|\Xi_{cc}~^2D_{\rho\rho}\frac{3}{2}^+\rangle$ is a fairly narrow
state with a width of a few MeV when its mass varies in the range.
Its strong decay is dominated by $\Lambda_cD$.

Taking the masses of $|\Xi_{cc}~^2D_{\rho\rho}\frac{5}{2}^+\rangle$
and $|\Xi_{cc}~^4D_{\rho\rho}\frac{5}{2}^+\rangle$ in the range of
(4.152-4.450) GeV, they are two narrow states with a width of
$\Gamma<4$ MeV and mainly decay into $\Lambda_cD$ if their masses
are below the threshold of $\Sigma^*_cD$. However, when the
$\Sigma^*_cD$ channel is open, the total decay widths of those two
states are sensitive to the mass and can increase up to several
tenths MeV. If so, their dominant decay modes should be
$\Sigma^*_cD$.

For the states $|\Xi_{cc}~^4D_{\rho\rho}\frac{1}{2}^+\rangle$ and
$|\Xi_{cc}~^4D_{\rho\rho}\frac{3}{2}^+\rangle$, if their masses are
above the threshold of $\Lambda_cD$, they mainly decay into
$\Lambda_cD$ and have a width of several tens MeV.

Taking the mass of $|\Xi_{cc}~^4D_{\rho\rho}\frac{7}{2}^+\rangle$ in
the range of (4.20 - 4.60) GeV, we get that the decay width of this
state is about $\Gamma\simeq(0-120)$ MeV. Its strong decays are
governed by the $\Lambda_cD$ channel in the whole mass region
considered in the present work. When we take the mass of
$|\Xi_{cc}~^4D_{\rho\rho}\frac{7}{2}^+\rangle$ with $M=4373$ MeV,
the predicted branching ratio is
\begin{equation}
\frac{\Gamma[\Lambda_cD]}{\Gamma_{\text{total}}}\simeq98\%.
\end{equation}
So, this state is most likely to be observed in the $\Lambda_cD$
channel.

Then, we analyze the decay properties of the $2D_{\rho\rho}$ states
in the $\Omega_{cc}$ family, and plot the partial decay widths and
total decay width as functions of the masses in Fig.~\ref{OMGDR}.

To investigate the decay properties of the
$|\Omega_{cc}~^2D_{\rho\rho}\frac{3}{2}^+\rangle$, we plot its decay
widths as a function of the mass in the range of $M=(4.34-4.45)$
GeV. From the figure, its strong decay width is around a few MeV.
This state mainly decays through the $\Xi_cD$ channel.

For the states $|\Omega_{cc}~^2D_{\rho\rho}\frac{5}{2}^+\rangle$ and
$|\Omega_{cc}~^4D_{\rho\rho}\frac{5}{2}^+\rangle$, we take their
masses in the range of $M=(4.34-4.60)$ GeV. If they lie below the
$\Sigma^*_cD$ threshold, the total decay widths are about $\Gamma<3$
MeV, and are dominated by $\Xi_cD$. However, if their masses are
above the threshold of $\Sigma^*_cD$, their dominant decay channels
should be $\Sigma^*_cD$ and their total decay widths may reach
several tenths MeV.

Taking the masses of
$|\Omega_{cc}~^4D_{\rho\rho}\frac{1}{2}^+\rangle$ and
$|\Omega_{cc}~^4D_{\rho\rho}\frac{3}{2}^+\rangle$ in the range of
$M=(4.34-4.40)$ GeV and $M=(4.34-4.50)$ GeV, respectively, their
decay widths depend considerably on their mass and are governed by
the $\Xi_cD$ channel.

Assuming the mass of the
$|\Omega_{cc}~^4D_{\rho\rho}\frac{7}{2}^+\rangle$ in the range of
(4.35-4.75) GeV, this state has a width of $\Gamma\simeq(0-150)$
MeV. If we take the mass of
$|\Omega_{cc}~^4D_{\rho\rho}\frac{7}{2}^+\rangle$ with $M=4523$ MeV,
the total decay width is about $\Gamma_{\text{total}}\simeq12$ MeV,
and the predicted branching ratio is
\begin{equation}
\frac{\Gamma[\Xi_cD]}{\Gamma_{\text{total}}}\simeq98\%.
\end{equation}

In brief, the $2D_{\rho\rho}$ states of $\Xi_{cc}$ and $\Omega_{cc}$
can decay through emitting a heavy-light meson when their masses are
above the threshold of $\Lambda_cD$ and $\Xi_cD$, respectively.
Their total decay widths maybe reach several tens MeV if their
masses are large enough. However, most of those states may lie below
the threshold of $\Lambda_cD$ or $\Xi_cD$, respectively.

\begin{figure*}[htpb]
\centering \epsfxsize=14.0 cm \epsfbox{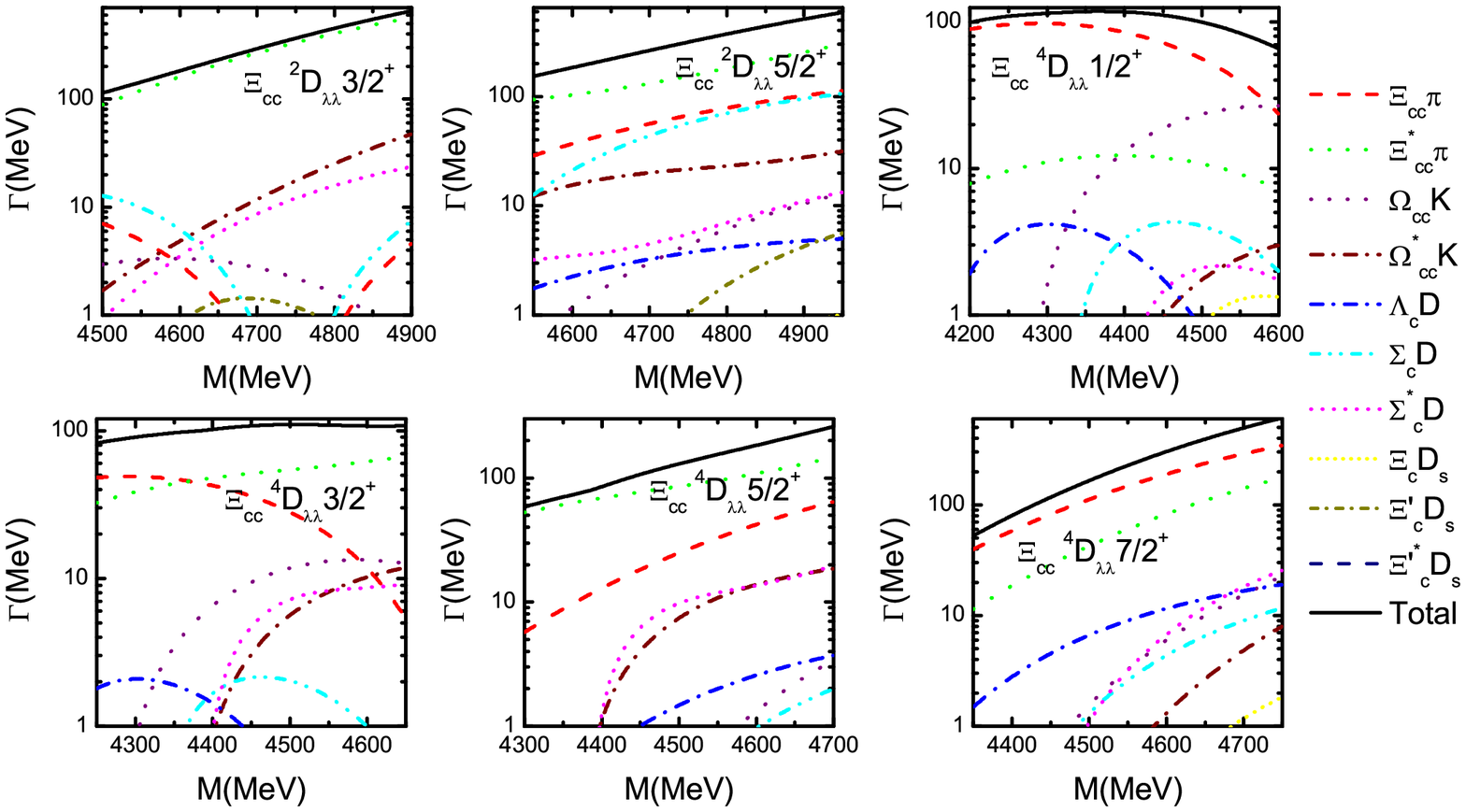} \caption{The
strong decay partial widths of the $2D_{\lambda\lambda}$-wave
$\Xi_{cc}$ states as a function of the mass.}\label{XiDL}
\end{figure*}

\begin{figure*}[htpb]
\centering \epsfxsize=14.0 cm \epsfbox{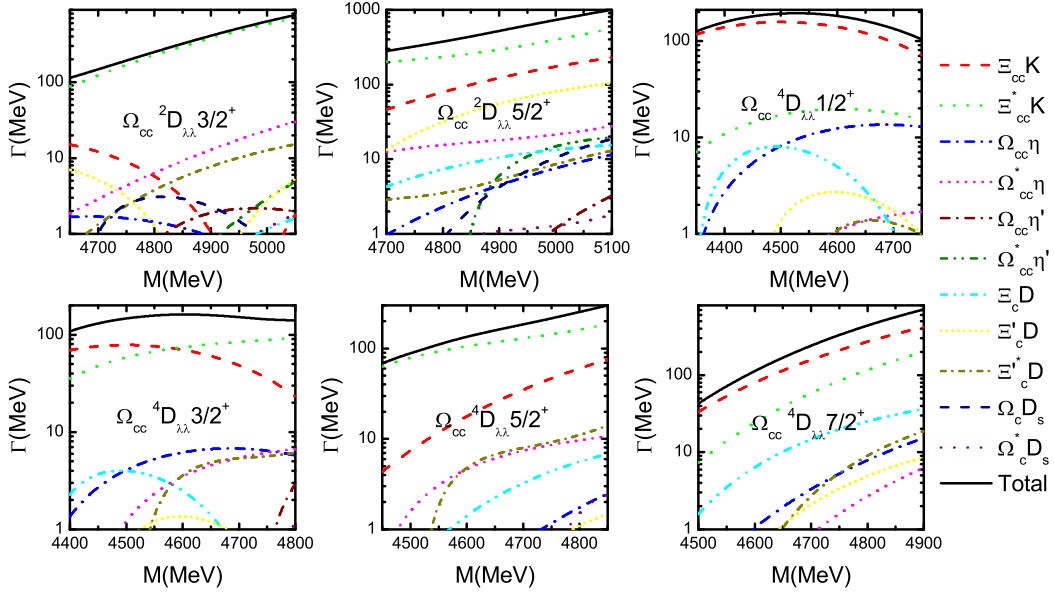} \caption{The
strong decay partial widths of the $2D_{\lambda\lambda}$-wave
$\Omega_{cc}$ states as a function of the mass.}\label{OMGDL}
\end{figure*}

\subsubsection{$\lambda$-mode excitations}

As emphasized in our previous work~\cite{Xiao:2017udy}, the
$\lambda$-mode orbitally excited state has relatively larger mass
than a $\rho$-mode orbitally excited state for the doubly charmed
baryons. The $2D_{\lambda\lambda}$ states should be heavier than the
$2D_{\rho\rho}$ states with the same $J^P$. Thus, many other decay
modes are allowed when we study the strong decay properties of
$2D_{\lambda\lambda}$ states.

In the $\Xi_{cc}$ family, we estimate the mass of the
$|\Xi_{cc}~^2D_{\lambda\lambda}\frac{3}{2}^+\rangle$ in the range of
(4.50-4.90) GeV, and then investigate its strong decay properties as
a function of the mass in Fig.~\ref{XiDL}. The decay width of the
state $|\Xi_{cc}~^2D_{\lambda\lambda}\frac{3}{2}^+\rangle$ is about
$\Gamma\simeq(100-650)$ MeV. The main decay channel is
$\Xi^*_{cc}\pi$ and the predicted branching ratio is
\begin{equation}
\frac{\Gamma[\Xi^*_{cc}\pi]}{\Gamma_{\text{total}}}\simeq(77-87)\%.
\end{equation}
On the other hand, the partial decay width of
$\Gamma[|\Xi_{cc}~^2D_{\lambda\lambda}\frac{3}{2}^+\rangle\to
\Sigma^*_cD]$ is sizable. The partial width ratio between
$\Sigma^*_cD$ and $\Xi^*_{cc}\pi$ is
\begin{equation}
\frac{\Gamma[|\Xi_{cc}~^2D_{\lambda\lambda}\frac{3}{2}^+\rangle\rightarrow
\Sigma^*_cD]}{\Gamma[|\Xi_{cc}~^2D_{\lambda\lambda}\frac{3}{2}^+\rangle\rightarrow
\Xi^*_{cc}\pi]}\simeq3.2\%
\end{equation}
when we fix the mass of this state on $M=4.70$ GeV.

For the state $|\Xi_{cc}~^2D_{\lambda\lambda}\frac{5}{2}^+\rangle$,
its mass might be in the range of (4.55-4.95) GeV. The dependence of
the strong decay properties of
$|\Xi_{cc}~^2D_{\lambda\lambda}\frac{5}{2}^+\rangle$ on the mass is
plotted in Fig.~\ref{XiDL} as well. According to the figure, we can
see that the state has a predicted width of $\Gamma\simeq(150-590)$
MeV, and mainly decays into $\Xi_{cc}\pi$ and $\Xi^*_{cc}\pi$. The
predicted partial width ratio is
\begin{equation}
\frac{\Gamma[|\Xi_{cc}~^2D_{\lambda\lambda}\frac{5}{2}^+\rangle\rightarrow
\Xi_{cc}\pi]}{\Gamma[|\Xi_{cc}~^2D_{\lambda\lambda}\frac{5}{2}^+\rangle\rightarrow
\Xi^*_{cc}\pi]}\simeq(30-38)\%.
\end{equation}
Meanwhile, the role of the $\Sigma_{c}D$ channel becomes more and more
important as the mass increases. The branching ratio is
\begin{equation}
\frac{\Gamma[\Sigma_{c}D]}{\Gamma_{\text{total}}}\simeq(8-18)\%.
\end{equation}

We estimate the mass of
$|\Xi_{cc}~^4D_{\lambda\lambda}\frac{1}{2}^+\rangle$ in the range of
(4.20-4.60) GeV and calculate its strong decay widths, which are
shown in Fig.~\ref{XiDL}. From the figure, the state
$|\Xi_{cc}~^4D_{\lambda\lambda}\frac{1}{2}^+\rangle$ is a moderate
state with a width of $\Gamma\simeq(65-118)$ MeV, and its strong
decays are governed by the $\Xi_{cc}\pi$ channel. The predicted
branching ratio is
\begin{equation}
\frac{\Gamma[\Xi_{cc}\pi]}{\Gamma_{\text{total}}}\simeq(36-91)\%.
\end{equation}
It should be pointed out that if the decay channel $\Omega_{cc}K$ is
opened, which is sensitive to the mass, the branching ratio of this
decay channel may reach ~41$\%$. Since the predicted width of
$|\Xi_{cc}~^4D_{\lambda\lambda}\frac{1}{2}^+\rangle$ is not broad,
this resonance might be observed in the $\Xi_{cc}\pi$ channel.

The mass of the state
$|\Xi_{cc}~^4D_{\lambda\lambda}\frac{3}{2}^+\rangle$ might be in the
range of (4.25-4.65) GeV, which is $\sim50$ MeV heavier than that of
the state  $|\Xi_{cc}~^4D_{\lambda\lambda}\frac{1}{2}^+\rangle$. We
plot the strong decay properties of
$|\Xi_{cc}~^4D_{\lambda\lambda}\frac{3}{2}^+\rangle$ as a function
of the mass in Fig.~\ref{XiDL}. The state
$|\Xi_{cc}~^4D_{\lambda\lambda}\frac{3}{2}^+\rangle$ may be a
moderate state with a width of $\Gamma\simeq(82-110)$ MeV, and its
strong decays are dominated by the $\Xi_{cc}\pi$ and $\Xi^*_{cc}\pi$
channels. However, the partial width of $\Xi_{cc}\pi$ decreases
dramatically with the mass. So, the predicted branching ratio of the
$\Xi_{cc}\pi$ channel varies in a wide range of
\begin{equation}
\frac{\Gamma[\Xi_{cc}\pi]}{\Gamma_{\text{total}}}\simeq(58-5)\%.
\end{equation}
The branching ratio of the $\Xi^*_{cc}\pi$ channel is stable, which
is
\begin{equation}
\frac{\Gamma[\Xi^*_{cc}\pi]}{\Gamma_{\text{total}}}\simeq(39-62)\%.
\end{equation}
This state has good potential to be discovered in the $\Xi_{cc}\pi$
and $\Xi^*_{cc}\pi$ channels.

For the state $|\Xi_{cc}~^4D_{\lambda\lambda}\frac{5}{2}^+\rangle$,
we plot its partial decay widths and total widths as a function of
the mass in the range of (4.30-4.70) GeV. From Fig.~\ref{XiDL}, its
total decay width is about $\Gamma\simeq(59-260)$ MeV. The partial
decay width ratio of the main two decay channels $\Xi_{cc}\pi$ and
$\Xi^*_{cc}\pi$ is
\begin{equation}
\frac{\Gamma[|\Xi_{cc}~^4D_{\lambda\lambda}\frac{5}{2}^+\rangle
\rightarrow
\Xi_{cc}\pi]}{\Gamma[|\Xi_{cc}~^4D_{\lambda\lambda}\frac{5}{2}^+\rangle
\rightarrow \Xi^*_{cc}\pi]}\simeq(11-44)\%.
\end{equation}

Meanwhile, from the Fig.~\ref{XiDL} we notice that the strong decays
of the state $|\Xi_{cc}~^4D_{\lambda\lambda}\frac{7}{2}^+\rangle$
are dominated by the $\Xi_{cc}\pi$ and $\Xi^*_{cc}\pi$ channels as
well, when the mass lies in the range of (4.35-4.75) GeV. But the
total decay width of
$|\Xi_{cc}~^4D_{\lambda\lambda}\frac{7}{2}^+\rangle$ is about
$\Gamma\simeq(52-610)$ MeV, which shows stronger dependency on the
mass than that of
$|\Xi_{cc}~^4D_{\lambda\lambda}\frac{5}{2}^+\rangle$, and the
predicted ratio between $\Xi_{cc}\pi$ and $\Xi^*_{cc}\pi$ is
\begin{equation}
\frac{\Gamma[|\Xi_{cc}~^4D_{\lambda\lambda}\frac{7}{2}^+\rangle
\rightarrow
\Xi_{cc}\pi]}{\Gamma[|\Xi_{cc}~^4D_{\lambda\lambda}\frac{7}{2}^+\rangle
\rightarrow \Xi^*_{cc}\pi]}\simeq(3.5-7.8).
\end{equation}

In addition, we extract the strong decays of the
$2D_{\lambda\lambda}$ states in the $\Omega_{cc}$ family, and plot
their decay properties as functions of the masses in
Fig.~\ref{OMGDL}. Usually, the mass of the $\Omega_{cc}$ resonances
is about 150 MeV larger than that of the $\Xi_{cc}$
resonances~\cite{Ebert:2002ig,Lu:2017meb}. Thus we estimate the mass
of the state $|\Omega_{cc}~^2D_{\lambda\lambda}\frac{3}{2}^+\rangle$
might be in the range of (4.65-5.05) GeV. According to our
theoretical calculations, $|\Omega_{cc}~^2D_{\lambda\lambda}\frac{3}{2}^+\rangle$ is a broad
state with a width of $\Gamma\simeq(114-769)$ MeV, and $\Xi^*_{cc}K$
almost saturates its total decay widths.

Meanwhile, the state
$|\Omega_{cc}~^2D_{\lambda\lambda}\frac{5}{2}^+\rangle$ is most
likely to be a very broad state as well, and the total decay width
is about $\Gamma\simeq(280-1000)$ MeV with the mass in the range of
(4.70-5.10) GeV. Its strong decays are governed by the $\Xi_{cc}K$
and $\Xi^*_{cc}K$ channels. The predicted partial width ratio
between $\Xi_{cc}K$ and $\Xi^*_{cc}K$ is
\begin{equation}
\frac{\Gamma[|\Omega_{cc}~^2D_{\lambda\lambda}\frac{5}{2}^+\rangle
\rightarrow
\Xi_{cc}K]}{\Gamma[|\Omega_{cc}~^2D_{\lambda\lambda}\frac{5}{2}^+\rangle
\rightarrow \Xi^*_{cc}K]}\simeq(23-41)\%.
\end{equation}
The partial width of $\Xi'_{cc}D$ is sizable as well. This broad
state might be hard to be observed in experiments.

Taking the mass of
$|\Omega_{cc}~^4D_{\lambda\lambda}\frac{1}{2}^+\rangle$ in the range
of (4.35-4.75) GeV, the state
$|\Omega_{cc}~^4D_{\lambda\lambda}\frac{1}{2}^+\rangle$ might be a
moderate state with a width of $\Gamma\simeq(104-194)$ MeV, and
mainly decays into the $\Xi_{cc}K$ channel. Such a moderate state
has some possibility to be observed in future experiments.

As to the state
$|\Omega_{cc}~^4D_{\lambda\lambda}\frac{3}{2}^+\rangle$, we plot its
strong decay properties as a function of the mass in the range of
(4.40-4.80) GeV in Fig.~\ref{OMGDL}. The total decay width of
$|\Omega_{cc}~^4D_{\lambda\lambda}\frac{3}{2}^+\rangle$ is
$\Gamma\simeq(108-161)$ MeV. Its strong decays are dominated by the
$\Xi_{cc}K$ and $\Xi^*_{cc}K$ channels, and the predicted partial
decay width ratio is
\begin{equation}
\frac{\Gamma[|\Omega_{cc}~^4D_{\lambda\lambda}\frac{3}{2}^+\rangle
\rightarrow
\Xi_{cc}K]}{\Gamma[|\Omega_{cc}~^4D_{\lambda\lambda}\frac{3}{2}^+\rangle
\rightarrow \Xi^*_{cc}K]}\simeq(1.95-0.24).
\end{equation}

The total decay width of
$|\Omega_{cc}~^4D_{\lambda\lambda}\frac{5}{2}^+\rangle$ is
$\Gamma\simeq(68-300)$ MeV with the mass in the range of (4.45-4.85)
GeV. From the Fig~\ref{OMGDL}, this state mainly decays through the
$\Xi^*_{cc}K$ channel. The branching ratio is
\begin{equation}
\frac{\Gamma[\Xi^*_{cc}K]}{\Gamma_{\text{total}}}\simeq(93-62)\%.
\end{equation}
The partial width of the $\Xi_{cc}K$ channel is sizable as well.

The partial decay widths of the
$|\Omega_{cc}~^4D_{\lambda\lambda}\frac{7}{2}^+\rangle$ strongly
depend on its mass. Taking the mass of
$|\Omega_{cc}~^4D_{\lambda\lambda}\frac{7}{2}^+\rangle$ in the range
of (4.50-4.90) GeV, the total decay width varies in a wide range of
$\Gamma\simeq(43-708)$ MeV. Its strong decays are governed by the
$\Xi_{cc}K$ and $\Xi^*_{cc}K$ channels, and the partial decay width
ratio is
\begin{equation}
\frac{\Gamma[|\Omega_{cc}~^4D_{\lambda\lambda}\frac{7}{2}^+\rangle
\rightarrow
\Xi^*_{cc}K]}{\Gamma[|\Omega_{cc}~^4D_{\lambda\lambda}\frac{7}{2}^+\rangle
\rightarrow \Xi_{cc}K]}\simeq(22-48)\%.
\end{equation}
Meanwhile, the partial decay width of $\Xi_cD$ is sizable, and the
predicted partial width ratio between $\Xi_cD$ and $\Xi_{cc}K$ is
\begin{equation}
\frac{\Gamma[|\Omega_{cc}~^4D_{\lambda\lambda}\frac{7}{2}^+\rangle
\rightarrow
\Xi_{c}D]}{\Gamma[|\Omega_{cc}~^4D_{\lambda\lambda}\frac{7}{2}^+\rangle
\rightarrow \Xi_{cc}K]}\simeq(4.7-8.6)\%.
\end{equation}

In conclusion, in the $\Xi_{cc}$ and $\Omega_{cc}$ family, the
$2D_{\lambda\lambda}$ states with $J^P=1/2^+$, $7/2^+$ mainly decay
into the ground state with $J^P=3/2^+$ through emitting a
light-flavor meson, while the $2D_{\lambda\lambda}$ states with
$J^P=3/2^+$, $5/2^+$ mainly decay into the ground state with
$J^P=1/2^+$ plus a light-flavor meson. The states
$|^4D_{\lambda\lambda}\frac{1}{2}^+\rangle$ and
$|^4D_{\lambda\lambda}\frac{3}{2}^+\rangle$ are most likely to be
the moderate states with the total widths of $\Gamma\sim100$ MeV,
which are insensitive to their masses, and might be discovered in
their dominant decay channels.

\section{Summary}\label{suma}

In the present work, we have systematically studied the strong decay
properties of the low-lying $1P$ and $2D$ doubly charmed baryons in
the framework of the $^3P_0$ quark pair creation model. Our main
results are summarized as follows.

For the $1P$ $\rho$-mode doubly charmed baryons, their decay widths
should be fairly narrow because of the absence of the strong decay
modes. In addition, for the $1P$ $\lambda$-mode excitations, the
states $|^2P_{\lambda}\frac{3}{2}^-\rangle$ and
$|^4P_{\lambda}\frac{3}{2}^-\rangle$ are predicted to be moderate
states with a width of $\Gamma\sim150$ MeV. Their strong decays are
governed by the $\Xi^*_{cc}\pi$ or $\Xi^*_{cc}K$ channel. However,
the states $|^2P_{\lambda}\frac{1}{2}^-\rangle$ and
$|^4P_{\lambda}\frac{5}{2}^-\rangle$ are most likely to be narrow
states with a total decay width of $\Gamma\sim40$ MeV, and their
strong decays are dominated by the $\Xi_{cc}\pi$ or $\Xi_{cc}K$
channel. Such narrow states have good potential to be observed in
future experiments. Meanwhile, the dominant decay mode of the state
$|^4P_{\lambda}\frac{1}{2}^-\rangle$ is $\Xi_{cc}\pi$ or $\Xi_{cc}K$
as well, but the total decay width of this state is about
$\Gamma>200$ MeV.

Since the strong decays of $2D_{\rho\rho}$ doubly charmed baryons
via emitting a light-flavor meson are forbidden, they mainly decay
via emitting a heavy-light meson with a total decay width of several
tens MeV if their masses are large enough. The partial strong decay
widths of the $2D_{\rho\rho}$ doubly charmed baryons strongly depend
on their masses. The measurement of masses in the future will be
helpful to understand their inner structures.

Within the range of mass we considered, the $2D_{\lambda\lambda}$
states with $J^P=1/2^+$, $7/2^+$ mainly decay through the
$\Xi^*_{cc}\pi$ or $\Xi^*_{cc}K$ channels, respectively, while the
$2D_{\lambda\lambda}$ states with $J^P=3/2^+$, $5/2^+$ mainly decay
through the $\Xi_{cc}\pi$ or $\Xi_{cc}K$ channels. It should be
remarked that the states $|^4D_{\lambda\lambda}\frac{1}{2}^+\rangle$
and $|^4D_{\lambda\lambda}\frac{3}{2}^+\rangle$ are most likely to
be discovered in their corresponding dominant decay channels because
of their not broad widths of $\Gamma\sim100$ MeV, which are
insensitive to their masses.

\section*{Acknowledgements }

We would like to thank Xian-Hui Zhong for very helpful discussions.
This work is supported by the National Natural Science Foundation of
China under Grants No.~11575008, No.~11705056, No.~11621131001 and
973 program. This work is also in part supported by China
Postdoctoral Science Foundation under Grant No.~2017M620492.

\begin{appendix}

\section{The decay mode with doubly charmed baryon plus a light-flavor meson}

The harmonic oscillator wave functions for the orbitally excited
baryons in our calculation are
\begin{eqnarray}
\psi(l_{\rho}, m_{\rho}, l_{\lambda}, m_{\lambda})&=&(-i)^l\Bigg
[\frac{2^{l_{\rho}+2}}{\sqrt{\pi}(2l_{\rho}+1)!!}\Bigg
]^{\frac{1}{2}}\Bigg (\frac{1}{\alpha_{\rho}}\Bigg
)^{\frac{3}{2}+l_{\rho}}\mathcal{Y}_{l_{\rho}}^{m_{\rho}}
(\textbf{p}_{\rho})  \nonumber \\
&&\times (-i)^l\Bigg
(\frac{2^{l_{\lambda}+2}}{\sqrt{\pi}(2l_{\lambda}+1)!!}\Bigg
)^{\frac{1}{2}}\Bigg (\frac{1}{\alpha_{\lambda}}\Bigg
)^{\frac{3}{2}+l_{\lambda}}\mathcal{Y}_{l_{\lambda}}^{m_{\lambda}}
(\textbf{p}_{\lambda})  \nonumber \\
&&\times \text{exp}\Bigg
(-\frac{\mathbf{p}^2_{\rho}}{2\alpha_{\rho}^2}-\frac{\mathbf{p}^2_{\lambda}}{2\alpha_{\lambda}^2}\Bigg
),
\end{eqnarray}
where
$\mathbf{p}_{\rho}=\frac{1}{\sqrt{2}}(\mathbf{p}_1-\mathbf{p}_2)$
and
$\mathbf{p}_{\lambda}=\frac{1}{\sqrt{6}}(\mathbf{p}_1+\mathbf{p}_2-2\mathbf{p}_3)$.

The ground state wave function of the meson is
\begin{equation}
\psi(0,0)=\Bigg (\frac{R^2}{\pi}\Bigg )^{\frac{3}{4}}\text{exp}\Bigg
(-\frac{R^2\mathbf{p^2_{ab}}}{2}\Bigg ),
\end{equation}
where $\mathbf{p_{ab}}$ stands for the relative momentum between the
quark and antiquark in a meson.

Since all the final states are in the $S$-wave ground states in the
present work, the momentum space integration
$I^{M_{L_A},m}_{M_{L_B},M_{L_C}}(\mathbf{p})$ can be further
expressed as $\Pi(l_{\rho A},m_{\rho A},l_{\lambda A},m_{\lambda
A},m)$. Based on Fig.~\ref{qkp}(a), the explicit form of the
momentum space integration $\Pi(l_{\rho A},m_{\rho A},l_{\lambda
A},m_{\lambda A},m)$ are presented in the following.

For the $S$-wave decay,
\begin{equation}
\Pi(0,0,0,0,0)=\beta |\mathbf{p}|\Delta_{0,0}.
\end{equation}

For the $P$-wave decay,
\begin{eqnarray}
\Pi(0,0,1,0,0)&=&\Bigg (\frac{1}{\sqrt{6}\lambda_2}-\frac{\lambda_3}{2\lambda_2}\beta|\mathbf{p}|^2 \Bigg )\Delta_{0,1},\\
\Pi(0,0,1,1,-1)&=&-\frac{1}{\sqrt{6}\lambda_2}\Delta_{0,1}\nonumber\\
               &=&\Pi(0,0,1,-1,1).
\end{eqnarray}

For the $D$-wave decay,
\begin{eqnarray}
\Pi(0,0,2,0,0)&=&\Bigg (\frac{\lambda_3^2}{2\lambda_2^2}\beta|\mathbf{p}|^3-\frac{\sqrt{6}\lambda_3}{3\lambda_2^2}|\mathbf{p}| \Bigg )\Delta_{0,2},\\
\Pi(0,0,2,1,-1)&=&\frac{\lambda_3}{\sqrt{2}\lambda_2^2}|\mathbf{p}|\Delta_{0,2}\nonumber\\
               &=&\Pi(0,0,2,-1,1).
\end{eqnarray}

Here,
\begin{eqnarray}
\lambda_1=\frac{1}{\alpha_{\rho}^2},~~~~\lambda_2=\frac{1}{\alpha_{\lambda}^2}+\frac{R^2}{3},~~~~\lambda_3=\frac{2}{\sqrt{6}\alpha_{\lambda}^2}+\frac{R^2}{\sqrt{6}},\\
\lambda_4=\frac{1}{3\alpha_{\lambda}^2}+\frac{R^2}{8},~~~~~~~~\beta=1-\frac{\lambda_3}{\sqrt{6}\lambda_2},
\end{eqnarray}
for the above expressions and
\begin{eqnarray}
\Delta_{0,0}&=&\Bigg(\frac{1}{\pi\alpha_{\rho}^2}\Bigg)^{\frac{3}{4}}\Bigg(\frac{1}{\pi\alpha_{\lambda}^2}\Bigg)^{\frac{3}{4}}\Bigg(\frac{R^2}{\pi}\Bigg)^{\frac{3}{4}}
\Bigg(\frac{\pi^2}{\lambda_1\lambda_2}\Bigg)
^{\frac{3}{2}}\text{exp}\Bigg[-\Bigg(\lambda_4-\frac{\lambda_3^2}{4\lambda_2}\Bigg)|\mathbf{p}|^2\Bigg]\nonumber\\
&&\times\Bigg(-\sqrt{\frac{3}{4\pi}}\Bigg)\Bigg(\frac{1}
{\pi\alpha_{\rho}^2}\Bigg)^{\frac{3}{4}}\Bigg(\frac{1}
{\pi\alpha_{\lambda}^2}\Bigg)^{\frac{3}{4}},\\
\Delta_{0,1}&=&\Bigg(\frac{1}{\pi\alpha_{\rho}^2}\Bigg)^{\frac{3}{4}}\Bigg(\frac{1}{\pi\alpha_{\lambda}^2}\Bigg)^{\frac{3}{4}}\Bigg(\frac{R^2}{\pi}\Bigg)^{\frac{3}{4}}
\Bigg(\frac{\pi^2}{\lambda_1\lambda_2}\Bigg)
^{\frac{3}{2}}\text{exp}\Bigg[-\Bigg(\lambda_4-\frac{\lambda_3^2}{4\lambda_2}\Bigg)|\mathbf{p}|^2\Bigg]\nonumber\\
&&\times\frac{3i}{4\pi}\Bigg(\frac{1}{\pi\alpha_{\rho}^2}\Bigg)^{\frac{3}{4}}
\Bigg(\frac{8}{3\sqrt{\pi}}\Bigg)^{\frac{1}{2}}\Bigg(\frac{1}{\alpha_{\lambda}^2}\Bigg)^{\frac{5}{4}},\\
\Delta_{0,2}&=&\Bigg(\frac{1}{\pi\alpha_{\rho}^2}\Bigg)^{\frac{3}{4}}\Bigg(\frac{1}{\pi\alpha_{\lambda}^2}\Bigg)^{\frac{3}{4}}\Bigg(\frac{R^2}{\pi}\Bigg)^{\frac{3}{4}}
\Bigg(\frac{\pi^2}{\lambda_1\lambda_2}\Bigg)
^{\frac{3}{2}}\text{exp}\Bigg[-\Bigg(\lambda_4-\frac{\lambda_3^2}{4\lambda_2}\Bigg)|\mathbf{p}|^2\Bigg]\nonumber\\
&&\times\frac{\sqrt{15}}{8\pi}\Bigg(\frac{1}{\pi\alpha_{\rho}^2}\Bigg)^{\frac{3}{4}}
\Bigg(\frac{16}{15\sqrt{\pi}}\Bigg)^{\frac{1}{2}}\Bigg(\frac{1}{\alpha_{\lambda}^2}\Bigg)^{\frac{7}{4}}.
\end{eqnarray}

\section{The decay mode with a singly heavy baryon plus a heavy-light meson}

From Fig.~\ref{qkp}(c), the momentum space integration $\Pi(l_{\rho
A},m_{\rho A},l_{\lambda A},m_{\lambda A},m)$ can be expressed in
the following.

For the $S$-wave decay,
\begin{equation}
\Pi(0,0,0,0,0)=\beta |\mathbf{p}|\Delta_{0,0}.
\end{equation}

For the $P$-wave decay,
\begin{eqnarray}
\Pi(0,0,1,0,0)&=&-\frac{1}{2f_1}\Bigg(f_2\beta|\mathbf{p}|^2+\zeta\Bigg)\Delta_{0,1},\\
\Pi(0,0,1,1,-1)&=&\Pi(0,0,1,-1,1)=\frac{\zeta}{2f_1}\Delta_{0,1},\\
\Pi(1,0,0,0,0)&=&\Bigg(\beta\varpi|\mathbf{p}|^2+\frac{1}{2\sqrt{2}\lambda_1}+\frac{\zeta\lambda_2}{4\lambda_1f_1}\Bigg)\Delta_{1,0},
\end{eqnarray}
\begin{eqnarray}
\Pi(1,1,0,0,-1)&=&-\Bigg(\frac{\zeta\lambda_2}{4\lambda_1f_1}+\frac{1}{2\sqrt{2}\lambda_1}\Bigg)\Delta_{1,0}\nonumber\\
               &=&\Pi(1,-1,0,0,1).
\end{eqnarray}

For the $D$-wave decay,
\begin{eqnarray}
\Pi(0,0,2,0,0)=\frac{f_2}{f_1^2}\Bigg(\frac{1}{2}\beta f_2|\mathbf{p}|^3+\zeta|\mathbf{p}|\Bigg)\Delta_{0,2},\\
\Pi(0,0,2,1,-1)=-\frac{\sqrt{3}f_2}{2f_1^2}\zeta|\mathbf{p}|\Delta_{0,2}
              =\Pi(0,0,2,-1,1), \\
\Pi(2,0,0,0,0)=2\Bigg(\beta\varpi^2|\mathbf{p}|^2+\frac{\varpi}{\sqrt{2}\lambda_1}
              +\frac{\lambda_2\zeta\varpi}{2\lambda_1f_1}\Bigg)|\mathbf{p}|\Delta_{2,0},
\end{eqnarray}
\begin{eqnarray}
\Pi(2,1,0,0,-1)&=&-\Bigg(\frac{\sqrt{6}\varpi}{2\lambda_1}+\frac{\sqrt{3}\lambda_2\varpi\zeta}{2\lambda_1f_1}\Bigg)|\mathbf{p}|\Delta_{2,0},\nonumber\\
               &=&\Pi(2,-1,0,0,1),\\
\Pi(1,0,1,0,0)&=&\Bigg(\frac{f_2\varpi\beta}{2f_1}|\mathbf{p}|^2+\frac{\lambda_2f_2\zeta}{8\lambda_1f_1^2}+\frac{\varpi\zeta}{2f_1}+\frac{\lambda_2\beta}{4\lambda_1f_1}\nonumber\\
                &&+\frac{\sqrt{2}f_2}{8\lambda_1f_1}\Bigg)|\mathbf{p}|\Delta_{1,1},
\end{eqnarray}
\begin{eqnarray}
\Pi(1,0,1,1,-1)=-\frac{\varpi\zeta}{2f_1}|\mathbf{p}|\Delta_{1,1}
=\Pi(1,0,1,-1,1),\\
\Pi(1,1,1,-1,0)=-\frac{\lambda_2\beta}{4\lambda_1f_1}|\mathbf{p}|\Delta_{1,1}
=\Pi(1,-1,1,1,0),
\end{eqnarray}
\begin{eqnarray}
\Pi(1,1,1,0,-1)&=&-\Bigg(\frac{\lambda_2f_2\zeta}{8\lambda_1f_1^2}+\frac{\sqrt{2}f_2}{8\lambda_1f_1}\Bigg)|\mathbf{p}|\Delta_{1,1}\nonumber\\
              &=&\Pi(1,-1,1,0,1),
\end{eqnarray}

Here,
\begin{eqnarray}
\lambda_1&=&\frac{1}{2\alpha_{\rho}^2}+\frac{1}{8\alpha'^2_{\rho}}+\frac{3}{8\alpha'^2_{\lambda}}+\frac{R^2}{4},\\
\lambda_2&=&-\frac{\sqrt{3}}{4\alpha'^2_{\rho}}+\frac{\sqrt{3}}{4\alpha'^2_{\lambda}}-\frac{\sqrt{3}R^2}{6},\\
\lambda_3&=&\frac{1}{2\alpha_{\lambda}^2}+\frac{3}{8\alpha'^2_{\rho}}+\frac{1}{8\alpha'^2_{\lambda}}+\frac{R^2}{12},\\
\lambda_4&=&\frac{\sqrt{2}}{4\alpha'^2_{\rho}}+\frac{\sqrt{2}}{4\alpha'^2_{\lambda}}+\frac{R^2}{\sqrt{2}}\frac{m_2}{m_2+m_5},\\
\lambda_5&=&-\frac{\sqrt{6}}{4\alpha'^2_{\rho}}+\frac{\sqrt{6}}{12\alpha'^2_{\lambda}}-\frac{R^2}{\sqrt{6}}\frac{m_2}{m_2+m_5},\\
\lambda_6&=&\frac{1}{4\alpha'^2_{\rho}}+\frac{1}{12\alpha'^2_{\lambda}}+\frac{R^2}{2}\Bigg(\frac{m_2}{m_2+m_5}\Bigg)^2,\\
      f_1&=&\lambda_3-\frac{\lambda_2^2}{4\lambda_1},\\
      f_2&=&\lambda_5-\frac{\lambda_2\lambda_4}{2\lambda_1},\\
      f_3&=&\lambda_6-\frac{\lambda_4^2}{4\lambda_1},
\end{eqnarray}
for the above expressions and
\begin{eqnarray}
\Delta_{0,0}&=&\Bigg(\frac{1}{\pi\alpha'^2_{\rho}}\Bigg)^{\frac{3}{4}}\Bigg(\frac{1}{\pi\alpha'^2_{\lambda}}\Bigg)^{\frac{3}{4}}\Bigg(\frac{R^2}{\pi}\Bigg)^{\frac{3}{4}}
\Bigg(\frac{\pi^2}{\lambda_1f_1}\Bigg)
^{\frac{3}{2}}\text{exp}\Bigg[-\Bigg(f_3-\frac{f_2^2}{4f_1}\Bigg)|\mathbf{p}|^2\Bigg]\nonumber\\
&&\times\Bigg(-\sqrt{\frac{3}{4\pi}}\Bigg)\Bigg(\frac{1}
{\pi\alpha_{\rho}^2}\Bigg)^{\frac{3}{4}}\Bigg(\frac{1}
{\pi\alpha_{\lambda}^2}\Bigg)^{\frac{3}{4}},\\
\Delta_{0,1}&=&\Bigg(\frac{1}{\pi\alpha'^2_{\rho}}\Bigg)^{\frac{3}{4}}\Bigg(\frac{1}{\pi\alpha'^2_{\lambda}}\Bigg)^{\frac{3}{4}}\Bigg(\frac{R^2}{\pi}\Bigg)^{\frac{3}{4}}
\Bigg(\frac{\pi^2}{\lambda_1f_1}\Bigg)
^{\frac{3}{2}}\text{exp}\Bigg[-\Bigg(f_3-\frac{f_2^2}{4f_1}\Bigg)|\mathbf{p}|^2\Bigg]\nonumber\\
&&\times\frac{3i}{4\pi}\Bigg(\frac{1}{\pi\alpha_{\rho}^2}\Bigg)^{\frac{3}{4}}
\Bigg(\frac{8}{3\sqrt{\pi}}\Bigg)^{\frac{1}{2}}\Bigg(\frac{1}{\alpha_{\lambda}^2}\Bigg)^{\frac{5}{4}},\\
\Delta_{1,0}&=&\Bigg(\frac{1}{\pi\alpha'^2_{\rho}}\Bigg)^{\frac{3}{4}}\Bigg(\frac{1}{\pi\alpha'^2_{\lambda}}\Bigg)^{\frac{3}{4}}\Bigg(\frac{R^2}{\pi}\Bigg)^{\frac{3}{4}}
\Bigg(\frac{\pi^2}{\lambda_1f_1}\Bigg)
^{\frac{3}{2}}\text{exp}\Bigg[-\Bigg(f_3-\frac{f_2^2}{4f_1}\Bigg)|\mathbf{p}|^2\Bigg]\nonumber\\
&&\times\frac{3i}{4\pi}\Bigg(\frac{8}{3\sqrt{\pi}}\Bigg)^{\frac{1}{2}}\Bigg(\frac{1}{\alpha_{\rho}^2}\Bigg)^{\frac{5}{4}}\Bigg(\frac{1}{\pi\alpha_{\lambda}^2}\Bigg)^{\frac{3}{4}},\\
\Delta_{0,2}&=&\Bigg(\frac{1}{\pi\alpha'^2_{\rho}}\Bigg)^{\frac{3}{4}}\Bigg(\frac{1}{\pi\alpha'^2_{\lambda}}\Bigg)^{\frac{3}{4}}\Bigg(\frac{R^2}{\pi}\Bigg)^{\frac{3}{4}}
\Bigg(\frac{\pi^2}{\lambda_1f_1}\Bigg)
^{\frac{3}{2}}\text{exp}\Bigg[-\Bigg(f_3-\frac{f_2^2}{4f_1}\Bigg)|\mathbf{p}|^2\Bigg]\nonumber\\
&&\times\frac{\sqrt{15}}{8\pi}\Bigg(\frac{1}{\pi\alpha_{\rho}^2}\Bigg)^{\frac{3}{4}}
\Bigg(\frac{16}{15\sqrt{\pi}}\Bigg)^{\frac{1}{2}}\Bigg(\frac{1}{\alpha_{\lambda}^2}\Bigg)^{\frac{7}{4}},\\
\Delta_{2,0}&=&\Bigg(\frac{1}{\pi\alpha'^2_{\rho}}\Bigg)^{\frac{3}{4}}\Bigg(\frac{1}{\pi\alpha'^2_{\lambda}}\Bigg)^{\frac{3}{4}}\Bigg(\frac{R^2}{\pi}\Bigg)^{\frac{3}{4}}
\Bigg(\frac{\pi^2}{\lambda_1f_1}\Bigg)
^{\frac{3}{2}}\text{exp}\Bigg[-\Bigg(f_3-\frac{f_2^2}{4f_1}\Bigg)|\mathbf{p}|^2\Bigg]\nonumber\\
&&\times\frac{\sqrt{15}}{8\pi}\Bigg(\frac{16}{15\sqrt{\pi}}\Bigg)^{\frac{1}{2}}\Bigg(\frac{1}{\alpha_{\rho}^2}\Bigg)^{\frac{7}{4}} \Bigg(\frac{1}{\pi\alpha_{\lambda}^2}\Bigg)^{\frac{3}{4}},\\
\Delta_{1,1}&=&\Bigg(\frac{1}{\pi\alpha'^2_{\rho}}\Bigg)^{\frac{3}{4}}\Bigg(\frac{1}{\pi\alpha'^2_{\lambda}}\Bigg)^{\frac{3}{4}}\Bigg(\frac{R^2}{\pi}\Bigg)^{\frac{3}{4}}
\Bigg(\frac{\pi^2}{\lambda_1f_1}\Bigg)
^{\frac{3}{2}}\text{exp}\Bigg[-\Bigg(f_3-\frac{f_2^2}{4f_1}\Bigg)|\mathbf{p}|^2\Bigg]\nonumber\\
&&\times\Bigg(-\frac{3}{4\pi}\Bigg)^{\frac{3}{2}}\frac{8}{3\sqrt{\pi}}\Bigg(\frac{1}{\alpha^2_{\rho}\alpha^2_{\lambda}}\Bigg)^{\frac{5}{4}}.
\end{eqnarray}
Here, the parameters $\alpha'_{\rho}$ and $\alpha'_{\lambda}$ stand
the harmonic oscillator parameters of the final singly heavy baryon.

\end{appendix}

\end{document}